\documentclass[%
 reprint,
 amsmath,amssymb,
 aps,
]{revtex4-2}
\usepackage[colorlinks=true,urlcolor=blue,linkcolor=blue,citecolor=blue]{hyperref}
\usepackage{amsmath} 
\usepackage{amssymb}
\usepackage{epsf}
\usepackage{color}
\usepackage{dcolumn}
\usepackage{bm}

\usepackage{graphicx}
\graphicspath{ {./images/} }

\begin{document}

\title{Spin-current model of electric polarization with the tensor gyromagnetic ratio}%

\author{Mariya Iv. Trukhanova$^{1,2}$}
\email{trukhanova@physics.msu.ru}
\author{Pavel A. Andreev$^{1}$}
\email{andreevpa@physics.msu.ru }

\affiliation{$^{1}$~Faculty of Physics,
Lomonosov Moscow State University, Leninskie Gory 1, 119991 Moscow, Russia  \vspace{1em} \\
$^{2}$~Theoretical Physics Laboratory, Nuclear Safety Institute, Russian Academy of Sciences,
B. Tulskaya 52, 115191 Moscow, Russia}

\begin{abstract}
A new spin-current model of electric polarization of spin origin is developed for magnetic structures with an anisotropic gyromagnetic-ratio tensor (g-factor). Three mechanisms of the magnetoelectric effect are proposed, arising from the symmetric Heisenberg exchange interaction, the Dzyaloshinskii–Moriya interaction, and spin–spin interactions associated with the odd anisotropy of the symmetric exchange between magnetic ions mediated by nonmagnetic ions. The dependence of the electric polarization on the spin density and the tensor g-factor is derived within the generalized spin-current model. New contributions to the macroscopic electric polarization arising in cycloidal and helicoidal spin orders and induced by the off-diagonal components of the gyromagnetic tensor are predicted. The extension of the spin-current model to include a tensor g-factor may be important for magnetic ferroelectrics containing heavy ions that contribute significantly to the magnetoelectric effect.
 
\end{abstract}
\maketitle

\section{Introduction}
Research on the magnetoelectric effect and multiferroic materials has advanced significantly in recent years \cite{KOPYL2021100149, Khomski}. At the same time, the focus of modern research remains on understanding the microscopic origin of the spin-driven magnetoelectric effect and developing its theoretical description \cite{Dong, Andreev}. Multiferroics are divided into two main groups \cite{Khomski2}, depending on the relationship between their magnetic and ferroelectric properties. The main example of type-I multiferroics is bismuth ferrite BiFeO$_3$, which has high critical magnetic and ferroelectric temperatures that differ by an order of magnitude \cite{PhysRevB.71.014113, PhysRevB.71.060401}. Thus, the magnetic and ferroelectric orders in such materials are only weakly coupled \cite{Catalan}. In type-II multiferroics, ferroelectricity is induced by magnetic ordering \cite{Kimura, Kimura2} and often appears in conjunction with a spiral magnetic phase, which can emerge in manganite oxide perovskites such as RMnO$_3$ (R = Tb, Dy, EuY) or Ni$_3$V$_2$O$_6$ \cite{Kimura}.

The basic principle for understanding strongly correlated magnetic dielectrics is the symmetry principle \cite{PhysRevLett.96.067601}. The necessary conditions for the existence of the linear magnetoelectric effect in materials are the separate breaking of time-reversal symmetry and space-inversion symmetry ($T$- and $P$-parity), while preserving the combined $PT$-parity \cite{Pyatakov}. However, to gain a clearer understanding of the physical principles underlying the magnetoelectric effect, various models and new conceptual approaches have been proposed. In accordance with Ref. \cite{Tokura}, two main mechanisms for the formation of spin-driven ferroelectricity have been developed to describe the origin of the electric dipole moment. The first mechanism arises from symmetric spin-exchange interactions in collinear magnetic structures without the necessary involvement of spin-orbit interaction \cite{PhysRevB.84.054440, PhysRevLett.102.026404}. The second mechanism produces polarization under the influence of spin-orbit coupling \cite{Tokura}. Y. Tokura explains that, in this case, the electric dipole moment arises in a cluster consisting of two magnetic ions and an oxygen ion located between them. The description of ferroelectricity in noncollinear spin orders within such a cluster was proposed by Katsura, Nagaosa, and Balatsky (KNB) \cite{PhysRevLett.95.057205} and is known as the spin-current model. In this model, the oxygen ion does not shift from the midpoint of the bond between two magnetic ions, and the microscopic electric dipole moment appears as a result of electronic charge-density deformation. On the other hand, an attempt to describe the magnetoelectric effect in spiral spin structures was made by I. Sergienko and E. Dagotto using the Dzyaloshinskii–Moriya interaction mechanism \cite{PhysRevB.73.094434}, in which the displacement of the oxygen ion from the midpoint leads to the formation of an electric dipole moment.

A new theoretical model based on the method of many-body quantum hydrodynamics has been developed and applied to substantiate the nature of electric polarization in various multiferroic spin structures \cite{10.1142/S0217979225500729}. The spin-current model of electric polarization has been justified and generalized to ferromagnetic and antiferromagnetic states using this method and the mean-field part of the spin–orbit interaction \cite{Andreev_2024, 10.1142/S0217979225500729, Trukhanova_2024}, where the displacement of the oxygen ion from the midpoint of the bond between magnetic ions and an arbitrary spin structure of the magnetic sublattice are taken into account. The direct relation between electric polarization and effective spin currents allows for the reconstruction of the structure of microscopic dipole moments caused by Heisenberg and Dzyaloshinskii–Moriya exchange interactions. The analysis presented in \cite{Andreev_2024} shows that considering the symmetric exchange interaction between ions in the form of the Heisenberg Hamiltonian leads to a structure of the electric dipole moment that is proportional to the vector product of spins and to an expression for the electric polarization that corresponds to the result of M. Mostovoy, up to a proportionality coefficient \cite{PhysRevLett.96.067601}. On the other hand, it is also shown that the antisymmetric anisotropic Dzyaloshinskii–Moriya exchange interaction leads to a structure of the electric dipole moment that is proportional to the scalar product of two neighboring spins. The displacement of oxygen ions produces a nonvanishing equilibrium spin current, leading to electric polarization even in the case of a spatially uniform spin-density distribution \cite{Andreev_2024}. These consequences of the generalized spin-current model contradict the description by Tokura. The spin-current model has also been applied to antiferromagnetic spin states in Ref. \cite{Trukhanova_2024}.

The magnetic ions of transition and rare-earth metals in perovskite oxides belong to the group of heavy ions, for which spin-orbit coupling and low crystal symmetry result in a tensor g-factor \cite{10.1063/1.3665634, condmat10020021}. This fact must be taken into account in the spin-current model and may lead to the prediction of new types of electric polarization structures. This article aims to extend the spin-current model to the case of a tensor g-factor using the method of many-particle quantum hydrodynamics.

This paper is organized as follows. In Sec. \eqref{SecI}, the formulation of the problem of including the tensor gyromagnetic ratio (throughout the remainder of this paper, we use the terminology "g-factor tensor") in the description of the spin-current model is presented. In Sec. \eqref{Sec2}, the derivation of the dependence of the electric polarization on the spin density and the tensor g-factor within the generalized spin-current model is presented. In Secs. \eqref{Sec3} and \eqref{Sec4}, the electric polarization for cycloidal and helicoidal spin orders caused by the symmetric exchange interaction and the Dzyaloshinskii–Moriya interaction is derived. In Sec. \eqref{Sec5}, a new type of electric polarization caused by a novel form of spin–spin interaction related to the odd anisotropy of the symmetric exchange interaction is derived. In Sec. \eqref{Sec6}, a brief summary of the obtained results is presented.

\section{Spin-current model for the tensor g-factor}\label{SecI}

\subsection{The formulation of the problem}
We employ the method of many-particle quantum hydrodynamics (MPQH) to describe nonequilibrium dynamics in systems consisting of a large number of interacting particles. This approach enables the derivation of a closed set of dynamical equations for the principal physical observables, including the particle-number density, spin density, electric polarization, and momentum density. The central idea of MPQH is to reformulate quantum mechanics in terms of the evolution of macroscopic fields whose dynamics are determined by the underlying many-particle wave function \cite{Andreev_2019, Andreev_2016, Andreev_2023}.

In the present work, we apply the MPQH formalism to develop a generalized spin-current model. As the microscopic building block, we consider a cluster composed of two positively charged magnetic transition-metal ions belonging to different magnetic sublattices, with arbitrary spin vectors $\bm{S}_i$ and $\bm{S}_j$, and a negatively charged diamagnetic oxygen ion $O^{2-}$ located between them. Unlike the original Katsura–Nagaosa–Balatsky model, the oxygen ion is allowed to be displaced from the midpoint of the bond connecting the magnetic ions.

In the present approach, the crystal is represented as an ensemble of elementary clusters, each constructed around the $i$-th magnetic ion. Each cluster consists of a magnetic transition-metal ion and its surrounding ligand environment responsible for the local electric polarization. Specifically, each cluster is constructed around a magnetic ion and includes one of its nearest magnetic neighbors and an oxygen ion. For example, the cluster centered on the $i$-th ion contains the neighboring $j$-th ion, whereas the next cluster, centered on the $j$-th ion, contains the neighboring $k$-th ion, and so on. Consequently, adjacent clusters share one magnetic ion, and every magnetic ion simultaneously belongs to two neighboring clusters. Within the many-particle quantum hydrodynamic description, each cluster is treated as an effective point-like quasiparticle, and its internal microscopic structure is not considered explicitly. Nevertheless, the electronic structure of each cluster and the microscopic interactions taking place within it—including crystal-field effects, spin-orbit coupling, and hybridization with the surrounding oxygen ions—determine the effective physical characteristics of the quasiparticle, such as its electric dipole moment $\bm{d}_i$, spin $\bm{s}_i$, gyromagnetic tensor $\gamma_{i}^{\alpha\beta}$, and other local parameters. Accordingly, the summation in the interaction Hamiltonian is performed over all clusters constituting the crystal. The transition from microscopic to macroscopic quantities is achieved by taking the quantum-mechanical average values of the corresponding microscopic operators $\hat{\bm{d}}_i$ and spin $\hat{\bm{s}}_i$ with respect to the many-particle wave function of the entire system. The magnetic lattice is decomposed into a sequence of overlapping nearest-neighbor clusters. 

We begin by specifying the interparticle interactions included in the model. We assume that each magnetic cluster acquires an induced electric dipole moment $\bm{d}_i$, that interacts with an external electric field. We further include the spin–orbit interaction and the exchange interactions governing the spin dynamics of the system. The evolution of the many-particle system is described by the Schrödinger–Pauli equation $i\hbar\partial_t\psi(R,t)=\hat{H}\psi(R,t)$ with the interaction Hamiltonian
\begin{align}
\hat{H}=-\sum_i^N\biggl(\hat{d}_i^\alpha E_i^\alpha+\gamma^{\alpha\nu}_i\hat{s}_i^{\nu} B^{\alpha}_i+\frac{\gamma_i^{\alpha\nu}}{m_ic}\varepsilon^{\alpha\beta\gamma}\hat{s}^{\nu}_iE^{\beta}_i\hat{p}_i^{\gamma}\biggr)\qquad\nonumber\\
-\frac{1}{2}\sum_{i,j\neq i}^{N}\biggl(U_{ij}(r_{ij})\hat{s}_i^\alpha\hat{s}_j^\alpha+D^\alpha_{ij}(r_{ij})\varepsilon^{\alpha\beta\gamma}\hat{s}^\beta_i\hat{s}^{\gamma}_j\biggr),\qquad\label{H}\end{align}
where $N$ is the total number of magnetic ions (particles) in the system, $\psi(R,t)$ is the many-particle wave function, which depends on the coordinates of all particles $R = {r_1, ...,r_N}$ and time $t$, $\hat{\bm{d}}_i$ is the operator of the electric dipole moment vector, $\hat{\bm{p}}_i=-i\hbar\hat{\bm{\nabla}}$ is the momentum operator of the $i$-th ion, and $\hbar$, $m_i$, and $c$ are the reduced Planck constant, the mass of the ions, and the speed of light in vacuum, respectively. $\bm{E}_i$ and $\bm{B}_i$ are the electric and magnetic field strengths acting on the $i$-th ion, $\hat{\bm{s}}_i$ is the dimensionless spin operator, which for the one-electron ions has the form of Pauli matrices, $\gamma_i^{\alpha\nu}$ is the gyromagnetic tensor, which is related to the tensor g-factor by the relation $\gamma^{\alpha\nu}=-\frac{|e|\hbar}{4m_ec}g^{\alpha\nu}=\gamma g^{\alpha\nu}$. Accordingly, the magnetic moment operator is defined as $\hat{\mu}^\alpha = \gamma g^{\alpha\beta}\hat{s}^\beta$, where $g_i^{\alpha\beta}$ is the dimensionless tensor g-factor. 

The first, second, and third terms represent the interaction energy of the electric dipoles with the external electric field, the Zeeman energy, and the spin–orbit interaction, respectively. The last two terms describe two different types of exchange interaction. The fourth term corresponds to the symmetric (Heisenberg) exchange interaction, where $U_{ij}(r_{ij})=U(\bm{r}_i-\bm{r}_j)$ is the scalar exchange integral. The fifth term represents the antisymmetric Dzyaloshinskii–Moriya exchange interaction, where $\bm{D}_{ij}(r_{ij})=\bm{D}(\bm{r}_i-\bm{r}_j)$ is the Dzyaloshinskii–Moriya interaction vector. All calculations throughout this paper are performed in Gaussian units.

Within the framework of many-particle quantum hydrodynamics, macroscopic observables are defined as the quantum-mechanical average values of the corresponding microscopic operators evaluated with respect to the many-particle wave function of the system. As an example, we consider the electric polarization. It is defined as the average value of the microscopic electric dipole density operator obtained by summing the dipole moment operators of all clusters, \begin{equation}\sum_i\delta(\bm{r} - \bm{r}_i)\hat{\bm{d}}_i\end{equation} in the form \begin{align}\label{P}
    P^\alpha(\bm{r},t)=\int \psi^{\dagger}(R,t)\sum_{i=1}^N\delta(\bm{r}-\bm{r}_i)\hat{d}^\alpha_i\psi(R,t)dR.
\end{align}In general, the electric dipole moment operator associated with an individual cluster possesses a nontrivial microscopic structure determined by its electronic configuration and the interactions within the cluster \cite{JETP}. Consequently, the macroscopic polarization reflects the underlying microscopic spin and crystal-field degrees of freedom through the effective electric dipole moment of each cluster.
 
The present study is closely related to several active directions of research in the physics of strongly correlated electron systems and multiferroic materials \cite{10.1093/nsr/nwz023}. One of these directions concerns the determination and theoretical description of local g-factor tensors in transition-metal and rare-earth compounds \cite{JENSEN197732}. In crystals, the local magnetic response of an ion is governed by the combined action of the crystal field, spin-orbit coupling, and covalent hybridization with the surrounding ligands. These interactions lift the degeneracy of the electronic states and produce an anisotropic g-factor tensor, whose principal values and orientation are determined by the local symmetry of the magnetic site rather than solely by the global crystallographic symmetry. Consequently, a g-factor tensor provides direct microscopic information on the local electronic structure \cite{Ganyushin2013} and magnetic anisotropy and has become one of the key quantities studied by electron paramagnetic resonance and inelastic neutron scattering.

A closely related research area is the crystal-field theory of transition-metal and rare-earth ions. The crystal field generated by the surrounding ligand environment splits the atomic multiplets and defines the low-energy magnetic states of the ion \cite{RUDOWICZ201528}. In transition-metal oxides, the interplay between the crystal field and spin-orbit coupling determines the magnetic anisotropy through the partial quenching of the orbital angular momentum and its subsequent restoration by spin-orbit interaction \cite{Podlesnyak_2021}. In rare-earth compounds, where the spin-orbit interaction is generally stronger than the crystal-field splitting, the crystal field acts within the ground-state multiplet and gives rise to highly anisotropic magnetic properties \cite{Sushkov2019} that are naturally described by tensor g-factors. The quantitative determination of crystal-field parameters \cite{Bonville2014} and g-factor tensors therefore represents an essential step in constructing microscopic models of magnetic interactions.

Magnetic anisotropy itself plays a fundamental role in determining the equilibrium spin structures of multiferroic materials. The competition between isotropic exchange interactions, the antisymmetric Dzyaloshinskii–Moriya interaction, magnetic anisotropy, and external magnetic fields stabilizes a variety of noncollinear magnetic states, including cycloidal, helicoidal, and conical spin textures \cite{Yang}. Since the anisotropy originates from spin-orbit coupling, its microscopic description inevitably requires a tensor relation between the magnetic moment and the spin degree of freedom. This observation motivates extending phenomenological and microscopic theories beyond the approximation of an isotropic scalar g-factor.

Another major research direction concerns the microscopic mechanisms of spin-induced ferroelectricity. Several complementary mechanisms have been proposed to explain the emergence of electric polarization in magnetically ordered systems. These include exchange striction associated with symmetric exchange interactions, the inverse Dzyaloshinskii–Moriya mechanism arising from spin-orbit coupling, spin-dependent hybridization between transition-metal and ligand orbitals, and purely electronic mechanisms described by the spin-current model. Although these approaches successfully explain many experimental observations, their common assumption of an isotropic magnetic response limits their applicability to systems containing heavy transition-metal or rare-earth ions, for which the magnetic moment is intrinsically anisotropic. Incorporating a tensor g-factor into microscopic theories of spin-induced polarization therefore provides a natural extension of existing models and offers the possibility of predicting additional contributions to the magnetoelectric coupling that are absent within the conventional isotropic approximation.

\subsection{The anisotropic g-factor of heavy ions}
It is well known that the magnetic moment is related to the mechanical moment through the gyromagnetic ratio.

Let us discuss the introduction of the gyromagnetic ratio tensor in more detail. In light atoms, the electrostatic interaction of electrons in the atomic shell structure significantly exceeds the spin-orbit interaction, and the so-called LS coupling is formed. In this case, on specific time scales, the effective magnetic moment and the total mechanical moment $\bm{J}_i$ of the $i$-th ion are approximately collinear, $\bm{\mu}_i=g_i\gamma\bm{J}_i$, when averaged over time. The Zeeman energy in the Hamiltonian is determined by the expression $H_{zeeman}=-\bm{\mu}_i\cdot\bm{B}_i=-g_i\gamma \bm{J}_i\cdot \bm{B}_i$, where $g_i$ is the Landé g-factor of the ion.

When the nuclear charge number in atoms exceeds 27, the energy of the residual electrostatic interaction becomes less than the energy of the spin-orbit interaction, and the LS coupling begins to break down, leading to a tensor nature of the gyromagnetic ratio. The breaking of the LS coupling primarily occurs in rare-earth ions, but we can also apply this approach to transition-metal ions. For most heavy transition-metal and rare-earth ions in multiferroics, the magnetic moment of the ion no longer lies along a straight line with the mechanical moment of the electron shell \cite{10.1063/5.0004125, PhysRevLett.103.156401}. In this case, we can represent the magnetic moment vector of the $i$-th ion as $\mu^\alpha_i=\gamma g^{\alpha\beta}_is^\beta_i$ \cite{PhysRevLett.109.197203}, where $g^{\alpha\beta}_i$ is the $3\times 3$ gyromagnetic tensor at the $i$-th site

\begin{equation} g^{\alpha\beta}=\begin{pmatrix} g^{xx} & g^{xy} & g^{xz} \\ g^{yx} & g^{yy} & g^{yz} \\ g^{zx} & g^{zy} & g^{zz} \end{pmatrix}. \end{equation}
Taking into account the tensor nature of the gyromagnetic ratio is important for describing the properties of multiferroics, such as rare-earth manganites with perovskite structures \cite{10.1063/1.3665634, Bonville2014}. In the spin-induced multiferroic TbMnO$_3$ \cite{condmat10020021, Kajimoto2011}, the g-factor tensor describes the anisotropic response of the magnetic moments of Tb$^{3+}$ and Mn$^{3+}$ ions to an external magnetic field, which originates from the orthorhombic crystal structure with the space group $P{bnm}$. It has been shown \cite{SOUMYA2023123971} that TbMnO$_3$ possesses a distorted perovskite structure, where Tb$^{3+}$ ions occupy the $A$ sites of the lattice, while Mn$^{3+}$ ions are located at the centers of oxygen octahedra (MnO$_6$).

The Mn$^{3+}$ ions are surrounded by oxygen octahedra formed by six oxygen ions. The Mn$^{3+}$ electronic configuration is $3d^4$, corresponding to the high-spin state, with three electrons occupying the lower-energy $t_{2g}$ orbitals and one electron occupying the higher-energy $e_g$ orbital in the octahedral crystal-field environment \cite{Mansouri2018, Kashchenko2017}. The Jahn–Teller distortion of the MnO$_6$ octahedra lifts the degeneracy of the $e_g$ orbitals and lowers the total electronic energy \cite{Dubrovskiy}. Such distortions, together with spin-orbit coupling, generate magnetic anisotropy and contribute to the anisotropic g-factor tensor of manganese ions, which may play an important role in the microscopic description of the magnetoelectric effect. The local symmetry of the Mn–O–Mn exchange bonds and the anisotropic magnetic environment can lead to additional contributions to the electric polarization beyond those predicted by the original isotropic spin-current model.

A multiferroic material may exhibit off-diagonal components of the g-factor tensor in the crystallographic coordinate system when the local symmetry of the magnetic-ion site does not coincide with the symmetry of the chosen reference frame. Such off-diagonal components are allowed by symmetry for magnetic sites with sufficiently low local symmetry and can also arise when the principal axes of the local crystal-field environment are rotated relative to the crystallographic axes. In addition, strong spin-orbit coupling enhances the anisotropy of the magnetic response, particularly in systems with partially unquenched orbital moments, including rare-earth ions such as Tb$^{3+}$ and Dy$^{3+}$, as well as transition-metal ions with significant orbital contributions. Although the g-factor tensor can always be diagonalized in the local principal-axis frame, its off-diagonal components in the crystallographic coordinate system represent physically meaningful quantities because they characterize the orientation of the local magnetic anisotropy relative to the crystal lattice.

The form of the g-factor tensor is completely determined by the crystal point group or, more precisely, by the local symmetry of the magnetic ion. For example, for orthorhombic symmetry, it has the form
\begin{equation}
g^{\alpha\beta}=\begin{pmatrix}
g^{x} & 0 & 0 \\ 0 & g^{y} & 0 \\ 0 & 0 & g^{z}
\end{pmatrix}.
\end{equation}
For tetragonal and trigonal symmetries, $g^x=g^y=g_{\perp}$ and $g^z=g_{\parallel}$. In both cases, all off-diagonal elements are forbidden.

The form of the g-factor tensor is determined by the local point symmetry of the magnetic ion and the crystallographic point group. In a monoclinic crystal system, the form of the g-factor tensor in the crystallographic basis $(a,b,c)$ is constrained by symmetry operations (typically a single twofold axis or a mirror plane, depending on the setting). For the most common case, where the unique axis is $\bm{b}$ (for example, for the space group $P2/c$), a single independent off-diagonal component $g^{xz}=g^{zx}$ is allowed. MnWO$_4$ and CuO are well-known multiferroics exhibiting spiral magnetic order \cite{PhysRevB.77.220407}. Their monoclinic crystal symmetry permits off-diagonal components of the g-factor tensor in the crystallographic frame. MnWO$_4$ is often regarded as a nearly uniaxial multiferroic \cite{Toledano_2010}, with $\bm b$ being the hard magnetic axis, whereas polarization components along $\bm a$ and $\bm c$ arise due to spin-orbit and magnetoelastic couplings.

Besides monoclinic systems, a tensor g-factor is also generally allowed in low-symmetry crystallographic environments, such as triclinic and orthorhombic systems with low-symmetry magnetic-ion sites. In the crystallographic coordinate system of triclinic crystals, no symmetry operations impose restrictions on the components of the g-factor tensor, and all components $g^{\alpha\beta}$ may, in principle, be nonzero.

In orthorhombic systems, the g-factor tensor can be diagonal in the principal-axis frame determined by the local crystal-field environment \cite{Khan2020, Zorko2011}. However, off-diagonal components may appear in the crystallographic coordinate system when the local principal axes of the magnetic ion are not aligned with the crystallographic axes or when the local site symmetry is reduced by structural distortions and anisotropic ligand environments. Representative examples include rare-earth orthoferrites $R$FeO$_3$ \cite{PRADHAN2025100082}, orthorhombic rare-earth manganites RMnO$_3$ (R = Tb, Dy, Eu) \cite{10.1093/nsr/nwz055}, and Jahn-Teller-active transition-metal oxides such as Cu-based compounds, where spin-orbit coupling and local crystal-field effects produce strongly anisotropic magnetic g-tensors. In all these systems, the experimentally observed g-factor tensor is determined by the local symmetry and electronic structure of the magnetic ion rather than by the global crystallographic symmetry alone.

Therefore, when describing magnetoelectric phenomena, we need to take into account that the $LS$ coupling is no longer valid and that the magnetic moments of transition-metal and rare-earth ions are generally not collinear with their mechanical moments. This feature of magnetoelectric materials should be taken into account in the development of the spin-current model. We incorporate the tensor nature of the gyromagnetic ratio (g-factor) into the many-particle interaction Hamiltonian \eqref{H}.

\section{The justification of the spin-current model}\label{Sec2}
\subsection{The momentum balance equation}
In this paper, we aim to establish the relation between the macroscopic electric polarization vector $P^\alpha$ and the tensor of the effective spin-current density $J_s^{\mu\nu}$ for a tensor gyromagnetic ratio, thereby generalizing the spin-current model.
Within the framework of the many-particle quantum hydrodynamics method, the spin-current model can be derived from the momentum balance equation and the corresponding force balance. It was shown in Ref. \cite{Andreev_2024} that the force balance in a crystal lattice leads to a relation between the equilibrium electric polarization and the spin-current density for the case of a scalar isotropic g-factor. To generalize this result to the case of a tensor g-factor, it is necessary to derive the corresponding momentum balance equation.
   
For this purpose, we introduce the microscopic definition of momentum density as the quantum-mechanical average of the momentum operator 
\begin{align}\label{p}
    p^\alpha(\bm{r},t)=\int \frac{1}{2}\biggl(\psi^{\dagger}(R,t)\sum_{i=1}^N\delta(\bm{r}-\bm{r}_i)\times\qquad\qquad\nonumber\\\times\hat{p}^\alpha_i\psi(R,t)+h.c.\biggr)dR,
\end{align}where $h.c.$ is used to denote the Hermitian conjugate term. The momentum density \eqref{p} in the vicinity of a point $\bm{r}$ in the physical space is determined as the quantum-mechanical average of the momentum density operator. We can further represent the microscopic density of all physical variables in the more compact form 
\begin{align}
    \xi(\bm{r},t) = \int dR \sum_i^N\delta(\bm{r}-\bm{r}_i)\psi^{\dagger}(R,t)\hat{\xi}_i\psi(R,t)\qquad\nonumber\\ = \biggl\langle\psi^{\dagger}\hat{\xi}_i\psi\biggr\rangle.\end{align}In statistical physics, averaging over time or over an ensemble is performed to replace an excessively complicated microscopic description with an effective macroscopic one. In quantum mechanics, however, averaging has a different meaning. To determine the most accurate value of a physical observable that can be assigned to a given quantum state, one evaluates the average value of the corresponding operator with respect to the wave function of that state. If the quantum state is an eigenstate of the operator associated with the observable, the average value is equal to the corresponding eigenvalue.

In the next step, we take the time derivative of the function \eqref{p} with respect to time and use the many-particle Pauli-Schrödinger equation with Hamiltonian \eqref{H}. This derivative acts on the wave function $\psi(R,t)$ in the integral allowing us to replace the partial time derivatives of the wave functions with the action of the Hamiltonian operator \begin{align}
    \partial_tp^\alpha=\biggl\langle\frac{i}{2\hbar}\biggl((\hat{H}\psi)^{\dagger}\hat{p}_i^\alpha \psi - \psi^{\dagger}\hat{p}^\alpha_i(\hat{H}\psi)\qquad\nonumber\\ -(\hat{p}_i^\alpha\psi)^\dagger (\hat{H}\psi)+ \hat{p}_i^{\alpha\dagger}(\hat{H}\psi)^{\dagger}\psi\biggr)\biggr\rangle.
\end{align}
For clarity, we present the derivation of the force density associated with the first term of the many-particle Hamiltonian \eqref{H}. The corresponding contribution to the momentum-balance equation illustrates the general procedure used to derive macroscopic force densities from the microscopic many-particle dynamics
\begin{align}
    \partial_tp^\alpha=\biggl\langle\frac{i}{2\hbar}\biggl(\psi^\dagger\biggl[\hat{p}_i^\alpha,\hat{d}_k^\nu E_k^\nu\biggr]\psi - \biggl(\biggl[\hat{p}_i^\alpha,\hat{d}_k^\nu E_k^\nu\biggr]\psi\biggr)^\dagger\psi\biggr)\biggr\rangle\nonumber\\= \biggl\langle \psi^\dagger(\nabla^\alpha_iE^\nu_k)\hat{d}_k^\nu\psi\biggr\rangle = \nabla^\alpha E^\nu \biggl\langle \psi^\dagger\hat{d}_i^\nu\psi\biggr\rangle = P^\nu \nabla^\alpha E^\nu,
\end{align} where, in deriving this term, we used the microscopic definition of the electric dipole moment density \eqref{P}. Owing to the presence of the operator $\delta(\bm r-\bm r_i)$, only the contributions with $i=k$ survive, corresponding to the interaction of the external electric field with the dipole moment of the same particle (cluster).

After the set of calculations, following the method of many-particle quantum hydrodynamics, we obtain the corresponding momentum balance equation
\begin{align}\label{momentum balance}
    \partial_tp^{\alpha}=\lambda_{ou}S^\beta\cdot(\partial^{\alpha}S^\beta)+P^\beta\cdot(\partial^{\alpha}E^\beta)\qquad\qquad\qquad\nonumber\\+\frac{\gamma^{\mu\nu}}{c}(\partial^{\alpha}E^{\beta})\varepsilon^{\beta\delta\mu}J^{\nu\delta}_s+\gamma^{\mu\nu}S^{\nu}(\partial^{\alpha}B^{\mu})+F^{\alpha}_{DM}.
\end{align} 
Let us discuss each term on the right-hand side of the momentum-balance equation \eqref{momentum balance}. The first force-density term on the right-hand side originates from the symmetric exchange interaction described by the Heisenberg Hamiltonian and is not related to relativistic spin-orbit coupling. Here,
$\lambda_{ou}=\int U(\xi)d\bm{\xi}$
is the exchange constant, which is determined by the integral characteristic of the interaction between neighboring magnetic ions. The second force-density term describes the action of a spatially nonuniform electric field on the electric dipole moments. The third term originates from spin-orbit coupling and is proportional to the product of the electric-field gradient and the spin-current density tensor $J_s^{\nu\delta}$. The fourth force-density term arises from the Zeeman interaction of the magnetic ions with the magnetic field. Its coupling strength is determined by the gyromagnetic tensor $\gamma^{\mu\nu}$. The last term represents the force density associated with the Dzyaloshinskii-Moriya interaction.

The microscopic definition of the spin-current tensor appearing in Eq. \eqref{momentum balance} can be derived in the form
\begin{equation}\label{spin current}
    J_s^{\alpha\beta}=\biggl\langle\frac{1}{2m_i} \biggl(\psi^{\dagger}(R,t)\hat{s}^{\alpha}_i\hat{p}^{\beta}_i\psi(R,t)+h.c.\biggr)\biggr\rangle.
\end{equation} In principle, the momentum-balance equation may also contain elastic restoring-force terms associated with the lattice potential. Since these terms are not proportional to the internal electric-field gradient, they do not affect the derivation of the constitutive relation between the electric polarization and the effective spin-current tensor and are therefore omitted in the present analysis.

In crystals, there are no particle flows, and the ions are localized in the nodes of the crystal lattice and oscillate around these nodes. However, we are interested in the description of the stationary state $\partial_t\bm{p}=0$, since it is important to obtain the relation between the spin current and the polarization. As we can see from Eq. \eqref{momentum balance}, this state is realized in the presence of spatially inhomogeneous fields that can be created by external sources, as well as electric dipoles and charges inside the crystal structure. Macroscopic equilibrium electric and magnetic fields exist in this state. All terms on the right-hand side contain spatial derivatives of electric and magnetic fields. For the interactions in the system, nonzero field values are possible in the presence of field inhomogeneity. And, as we can see, the second and third force densities on the right-hand side are proportional to the spatial derivative of the electric field $\partial^\beta E^\alpha$. This allows for the formation of a balance of these forces in the equilibrium state for various types of electric field inhomogeneity.
This leads to the formation of equilibrium polarization, which is defined by the effective spin current tensor, due to the spin-orbit coupling
\begin{equation}\label{PSC}
    P^{\alpha}(\bm{r},t)=\frac{\gamma^{\mu\nu}}{c}\varepsilon^{\alpha\mu\delta}J^{\nu\delta}_s(\bm{r},t).
\end{equation}This dependence establishes the relation between the macroscopic electric polarization and the spin-current density as a result of spin-orbit coupling in the internal electric field of the crystal lattice.

In our model, the electric field is the microscopic internal field of the crystal lattice generated by ionic charges and electric dipoles. This field is generally spatially nonuniform on the interatomic scale. The derivation of Eq. (8) does not require a particular form or symmetry of the tensor $\partial^\alpha E^\beta$. We emphasize that the spatial derivative of the electric field is a second-rank tensor whose microscopic form is determined by the internal electrostatic field of the crystal. In general, this tensor is arbitrary and may, if necessary, be decomposed into symmetric and antisymmetric parts. However, no specific symmetry of $\partial^\alpha E^\beta$ is required in the present derivation.

The force-balance condition is formulated locally, i.e., at each point of the crystal. Consequently, the same local electric-field-gradient tensor enters both the electrostatic force acting on the electric dipoles and the spin-orbit force. Since these force densities contain the identical tensor factor $\partial^\alpha E^\beta$, the local equilibrium condition allows this common factor to be eliminated without imposing additional assumptions on its microscopic structure.

Naturally, the electric-field gradient varies throughout the crystal and may change substantially near defects, interfaces, or sample boundaries. Nevertheless, the generalized spin-current model does not require an explicit description of these microscopic variations. Instead, the microscopic electric field enters only as a mediator of the local spin-orbit coupling, while the resulting constitutive relation between the electric polarization and the effective spin-current tensor is independent of the detailed spatial structure of the internal electric field.

As will be shown below, the effective spin currents arise from different types of interactions between the lattice ions. The expression for the electric polarization given by Eq. \eqref{PSC} generalizes the spin-current model and describes the orientation of the induced electric polarization for various incommensurate magnetic states in the presence of a tensor g-factor.

On the other hand, the combination of the first, fourth, and fifth terms on the right-hand side of the momentum-balance equation \eqref{momentum balance} yields an equilibrium magnetic field proportional to the spin density and the spatial derivative of the spin density, arising from the Heisenberg exchange interaction and the Dzyaloshinskii-Moriya interaction.

The Dzyaloshinskii-Moriya interaction force density for an odd Dzyaloshinskii coefficient has a structure distinct from the other terms in the momentum balance equation. We may represent this force field density in the form
\begin{align}
    F^\alpha_{DM}=\lambda_{\beta}\biggl((\bm{\delta}\cdot\bm{S})\nabla^\alpha(\bm{\nabla\cdot\bm{S}}) - (\bm{S}\cdot\bm{\nabla})\nabla^\alpha(\bm{\delta}\cdot\bm{S})\biggr)\nonumber\\=\lambda_{\beta}S^\mu\nabla^\alpha\biggl(\delta^\mu (\bm{\nabla\cdot\bm{S}}) - \bm{\delta}\cdot\nabla^\mu\bm{S}\biggr),\qquad
\end{align} and extract the equilibrium magnetic field from the equation
\begin{align}
    \lambda_{ou}S^\mu+\gamma^{\nu\mu}B^\nu+\lambda_{\beta}\biggl(\delta^\mu(\bm{\nabla}\cdot\bm{S}) -\bm{\delta}\cdot\nabla^\mu\bm{S}\biggr) = 0.
\end{align}However, since the expression for the force caused by the Dzyaloshinskii-Moriya interaction contains two spatial derivatives, it has a relatively small contribution compared to the other terms.

\subsection{Derivation of the structure of spin currents }
After obtaining the relation between the electric polarization and the spin-current density tensor \eqref{PSC}, it is necessary to reconstruct the dependence of the macroscopic electric polarization on the spin density and to relate this dependence to the type of interparticle interaction responsible for its formation. To this end, we derive the spin-density balance equation and determine the corresponding structures of the effective spin currents. We introduce the spin-density vector in the vicinity of a point in physical space as the quantum-mechanical average of the spin-density operator $\sum_i \delta(\bm{r}-\bm{r}_i)\hat{s}^{\alpha}_i$
\begin{align}\label{spin evolution}
S^\alpha(\bm{r},t)=\int dR \sum_i^N\delta(\bm{r}-\bm{r}_i)
\psi^{\dagger}(R,t)\hat{s}^\alpha_i\psi(R,t).
\end{align}
The time derivative acts on the wave function $\psi(R,t)$, and the time derivative of the wave function is replaced by the Hamiltonian \eqref{H} according to the Pauli equation,
\begin{align}
    \partial_tS^\alpha=\frac{i}{\hbar}\int\sum_i^N\delta(\bm{r}-\bm{r}_i)\psi^{\dagger}(R,t)[\hat{H},\hat{s}_i^\alpha]\psi(R,t)dR.
\end{align}
After a series of straightforward calculations, the spin-density evolution equation is obtained in the form
\begin{align}\label{spin evolution}
\partial_tS^\alpha+\partial\mu J_s^{\alpha\mu}=\frac{2\gamma^{\mu\beta}}{\hbar}\varepsilon^{\alpha\mu\nu}S^\beta B^\nu \qquad\qquad\qquad\nonumber\\
+\varepsilon^{\alpha\mu\nu}\lambda_u S^\mu\triangle S^\nu +T^\alpha_{DM},
\end{align}
where the Dzyaloshinskii–Moriya torque density depends on the spin density and can be written as
\begin{equation}
T^\alpha_{DM}=\lambda_{\beta}\delta^\nu\biggl(\varepsilon^{\mu\lambda\nu}S^\mu\partial^\lambda S^\alpha - \varepsilon^{\alpha\lambda\nu}S^\beta\partial^\lambda S^\beta\biggr),
\end{equation}
where we introduce the exchange-interaction coefficient
$\lambda_{\beta}=\frac{1}{3}\int \xi^2 \beta(\xi)d\bm{\xi}$.

The microscopic mechanism of this interaction can be described as follows. The indirect exchange interaction between magnetic ions mediated by a nonmagnetic oxygen ion gives rise to a spin-orbit contribution to the exchange energy. When neighboring magnetic moments are noncollinear, a lattice distortion accompanied by a displacement of the oxygen ion modifies the local symmetry of the exchange pathway between the magnetic ions. Such a displacement breaks the inversion symmetry of the bond and allows for the appearance of a nonzero Dzyaloshinskii-Moriya vector $\bm{D}_{ij}$. In multiferroic rare-earth manganite perovskite oxides, transition-metal magnetic ions interact indirectly through intermediate oxygen ions via superexchange processes involving hybridization between oxygen $p$ orbitals and transition-metal $d$-orbitals. This mechanism leads to the following form of the Dzyaloshinskii vector:

\begin{equation}\label{D}
D^\alpha_{ij}(r_{ij})=\beta(r_{ij})\varepsilon^{\alpha\mu\nu}r^\mu_{ij}\delta^\nu,
\end{equation}
where the proportionality coefficient $\beta$ depends on the relative distance between the magnetic ions (or clusters), $\bm{r}_{ij}$, and decreases rapidly with increasing distance, so that the interaction is effectively limited to nearest-neighbor ions. Throughout this work, we assume that the oxygen ion in each three-ion cluster is displaced by the same vector, $\bm{\delta}_i=\bm{\delta}$.
 
We can express the torques generated by the exchange interactions in terms of the spatial divergence of the spin-current density tensor by rewriting the spin-density balance equation in the form
$\partial_tS^\alpha+\partial_\mu J^{\alpha\mu}_s=0.$
Applying this procedure to the antisymmetric Dzyaloshinskii–Moriya interaction, we separate the divergence of the spin-current density tensor $J^{\alpha\mu}{DM}$ from the torque expression,
\begin{equation}
T^\alpha_{DM}= - \partial_\mu J^{\alpha\mu}_{DM} + \lambda_{\beta}\delta^\nu \varepsilon^{\mu\lambda\nu}S^\mu\partial^\lambda S^\alpha,
\end{equation}
where the spin-current density tensor is given by
\begin{equation}\label{JDM}
J^{\alpha\mu}_{DM}=\frac{\lambda_{\beta}}{2}\delta^\nu \varepsilon^{\alpha\mu\nu} \bm{S}^2.
\end{equation}
As follows from Eq. \eqref{JDM}, this effective spin current reaches its maximum value for collinear spin configurations and exists even for a spatially uniform spin-density distribution, being proportional to $\bm{S}^2$. Furthermore, the spin-current density is proportional to the displacement $\delta^\nu$ of the nonmagnetic oxygen ion from the midpoint between two magnetic ions in each cluster.

The Dzyaloshinskii–Moriya torque is thus decomposed into the divergence of an effective spin-current density tensor and a residual local torque term. The residual term is not assumed to be small. Rather, it represents a local torque acting on the spin density that cannot be expressed as the divergence of a spin-current tensor. Consequently, it contributes to the local spin dynamics but does not enter the definition of the effective spin current used in the present work. In the present context, the effective spin current should not be interpreted as a real transport of spin angular momentum. Rather, it emerges from the divergence of the torque density associated with different exchange interactions. We note that the definition of the spin current is not unique, since an arbitrary curl of a vector function can, in principle, be added without changing its divergence. Such a transformation, however, corresponds to a redefinition of the spin current itself. In the present work, we do not aim to introduce additional complexity into the model. Additional contributions that are not included in the effective Dzyaloshinskii–Moriya spin current considered here can be incorporated in future extensions of the theory.

The first term on the right-hand side of Eq. \eqref{spin evolution} represents the magnetic-field torque density and corresponds to the torque term appearing in the conventional Landau–Lifshitz equation. The second torque-density term originates from the symmetric exchange interaction described by the Heisenberg Hamiltonian, where $\lambda_u=\frac{1}{6}\int \xi^2 U(\xi) d\bm{\xi}$ is the second exchange constant, corresponding to the second moment of the exchange integral, whereas the constant $\lambda_{ou}$ may be regarded as the zeroth moment of the exchange integral. Accordingly, the second term on the right-hand side of Eq. \eqref{spin evolution} can be represented as the divergence of the spin-current tensor,
\begin{align}\label{HSC}
\lambda_u\varepsilon^{\alpha\mu\nu}S^\mu\triangle S^\nu = \partial_\delta\biggl(\lambda_u\varepsilon^{\alpha\mu\nu}S^\mu\partial^\delta S^\nu\biggr), \qquad\nonumber\\
J^{\alpha\delta}_{ex} = - \lambda_u\varepsilon^{\alpha\mu\nu}S^\mu\partial^\delta S^\nu.
\end{align}

We have thus established the generalized spin-current model and obtained a general expression for the macroscopic electric polarization arising from different types of interactions between the magnetic ions.

\subsection{Tensor g-factor in electric polarization}
Substituting the spin current form \eqref{HSC} into the polarization \eqref{PSC}, we obtain the macroscopic dependence of the polarization on the spin density
\begin{equation}\label{PHH}
    P^\alpha = - \frac{\lambda_u}{c}\gamma^{\mu\nu}\varepsilon^{\alpha\mu\delta}\varepsilon^{\nu\beta\lambda} S^\beta\partial^\delta S^\lambda.
\end{equation} 
It should be clarified that the mechanisms that transform spin current into polarization and the interaction that generates the spin structure are distinct. Spin-orbit interaction is a relativistic mechanism that leads to the transformation of spin current into electric polarization. In this case, the spin-orbit coupling is formed in a crystallographic field acting on magnetic ions. On the other hand, the Dzyaloshinskii-Moriya interaction generates a noncollinear spin structure. In turn, the antisymmetric anisotropic indirect exchange interaction—or Dzyaloshinskii-Moriya interaction is based on the contribution of the quantum effect of electron density overlap to the spin-orbit interaction energy. But, the electric polarization arising against the background of a noncollinear spin structure is characterized by the spin current which is caused by the symmetric Heisenberg exchange interaction.

For the case of anisotropic tensor g-factor (gyromagnetic ratio), we can generalize the expression for the macroscopic polarization \eqref{PDMiso} using \eqref{PSC} and \eqref{JDM}   
\begin{align}\label{PDM}
    P^\alpha = \frac{\gamma^{\mu\nu}}{c}\varepsilon^{\alpha\mu\delta}J^{\nu\delta}_{DM}\qquad\qquad
 \qquad\qquad\nonumber\\ =  \frac{\lambda_{\beta}}{2c}\bm{S}^2\biggl(\gamma^{\mu\mu}\delta^\alpha - \gamma^{\mu\alpha}\delta^\mu\biggr).\end{align} And, as we can see from the definition \eqref{PDM}, the direction of electric polarization  is independent of the orientation of the spin spiral and is determined by the displacement vector $\delta^\alpha$ of the diamagnetic oxygen ions from the position of the center of mass of the bond of magnetic ions.
 
The effective spin current associated with exchange interactions does not represent a physical transport of particles or spin angular momentum through the crystal. Instead, it is a field-induced quantity determined by the spatial variation of the spin density generated by the exchange interactions. Within the microscopic spin-balance equation, it is defined as the quantity whose divergence reproduces the exchange contribution to the spin dynamics.

The main objective of this paper is not merely to replace the scalar gyromagnetic ratio by a tensor quantity, but to develop a microscopic generalization of the spin-current model in which the local tensor g-factor is incorporated directly into the many-particle Hamiltonian. This approach leads to new tensorial relations between the electric polarization and the effective spin current and predicts additional polarization components that are absent in the conventional isotropic theory. The off-diagonal components of the g-factor tensor can generate additional macroscopic polarization, but the mscroscopic contribution depends on the magnetic point group and summation over all magnetic sublattices.

\section{Electric polarization caused by the symmetric exchange interaction}\label{Sec3}
\subsection{The cycloidal spin structure}

\subsubsection{The isotropic case}
According to the generalized spin-current model for the isotropic scalar g-factor \cite{Andreev_2024, 10.1142/S0217979225500729}, the dependence of the electric polarization on the spin density can be derived in the form
\begin{align}\label{IsotropicP}
P^\alpha=\frac{\gamma}{c}\lambda_u\biggl((\bm{S}\cdot\bm{\nabla})S^\alpha - S^\alpha (\bm{\nabla}\cdot\bm{S})\biggr).
\end{align}
Equation \eqref{IsotropicP} reproduces the phenomenological expression for the macroscopic polarization induced by incommensurate spin-density-wave states derived in Ref. \cite{PhysRevLett.96.067601}, up to an overall proportionality coefficient. For a cycloidal spin structure in the $xy$ plane, we introduce the wave vector $\bm{q}=q\bm{e}x$ and the spin-density vector
\begin{equation}\label{spin helix}
\bm{S}=S_{ox} \cos(qx)\bm{e}_x+S_{oy} \sin(qx)\bm{e}_y,
\end{equation}
where the $x$, $y$, and $z$ axes are identified with the crystallographic $b$, $c$, and $a$ axes, respectively. Equation \eqref{IsotropicP} then yields the following expression for the macroscopic electric polarization:
\begin{align}\label{IsotropicPy}
\bm{P}=-\frac{\gamma}{c}\lambda_uS_{ox}S_{oy}(\bm{e}_z\times\bm{q}).
\end{align}
In the isotropic case, the electric polarization lies in the spiral plane and is directed along the crystallographic $c$ axis, in agreement with the experimental observations. 

\subsubsection{The tensor g-factor}
The first mechanism that can be considered within the framework of the generalized spin-current model is associated with the symmetric exchange interaction and is described by the Heisenberg Hamiltonian.

In real multiferroic materials, the spin state can undergo a transition at temperatures $T < T_H$ to a cycloidal spin structure confined to a plane in the absence of an external magnetic field, thereby inducing ferroelectricity. Such a spin structure is observed in a number of perovskite oxides. For an arbitrary gyromagnetic-ratio tensor, the corresponding electric polarization is characterized by two nonzero components
\begin{align}\label{projection of Py}
    P^y=\frac{\gamma^{zz}}{c}\lambda_u\biggl(S^y\partial_xS^x - S^x\partial_x S^y\biggr)\qquad\nonumber\\=-\frac{\gamma^{zz}}{c}\lambda_uqS_{ox}S_{oy},\qquad\end{align} 
    \begin{align}\label{projection of Pz}
        P^z=  \frac{\gamma^{yz}}{c}\lambda_u\biggl(S^x\partial_xS^y - S^y\partial_x S^x\biggr)\qquad\nonumber\\=\frac{\gamma^{yz}}{c}\lambda_uqS_{ox}S_{oy}, \qquad
    \\ P^x = 0.\qquad\qquad\qquad\end{align}
\begin{figure}
\includegraphics[width=8.4cm,angle=0]{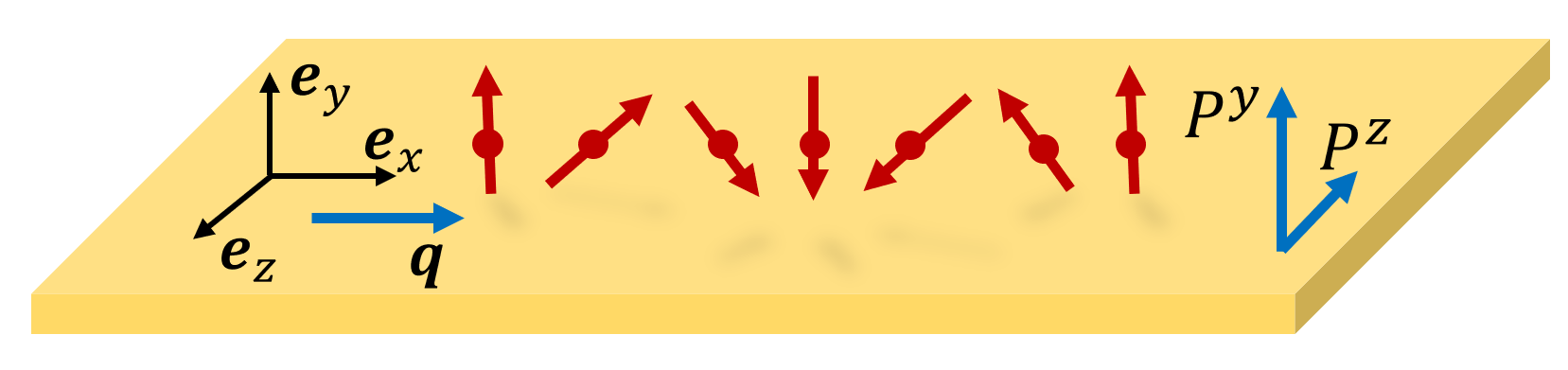}
\caption{\label{fig1} (Color online)
The cycloid spiral spin structure in the absence of an external magnetic field \eqref{spin helix}. The cycloidal structure lies in the $xy$-plane with the wave vector $\bm{q}=(q, 0, 0)$ which lies in this plane. Two projections of electric polarization \eqref{projection of Py} and \eqref{projection of Pz} appear  according to the definition \eqref{PHH}. One of them is determined by the non-diagonal component of g-factor.}
\end{figure}
The microscopic interpretation of this electric polarization is based on the idea that an effective spin current is generated by a noncollinear spin state through the symmetric Heisenberg exchange interaction, which does not require an explicit displacement of the oxygen-ion sublattice. In contrast, the Dzyaloshinskii-Moriya interaction is associated with a reduction of the local symmetry of the exchange pathway and can be accompanied by ligand displacements that stabilize a canted spin structure. The Dzyaloshinskii-Moriya interaction provides an important mechanism for the emergence of noncollinear magnetic order by lifting the degeneracy of collinear spin configurations and favoring a finite canting between neighboring magnetic moments. The electric polarization given by Eq. \eqref{projection of Py} is proportional to the symmetric Heisenberg exchange constant $\lambda_u$ and generalizes the isotropic result \eqref{IsotropicPy}, where the scalar gyromagnetic ratio is replaced by the tensor component $\gamma^{zz}$ (see Fig. \eqref{fig1}).

In addition to the $y$ component of the electric polarization, the anisotropy of the g-factor tensor gives rise to an additional nonzero component of the electric polarization along the $z$ direction, even in the absence of an external magnetic field. As follows from Eq. \eqref{projection of Pz}, this contribution is determined by the off-diagonal component of the gyromagnetic-ratio tensor $\gamma^{yz}$.

According to the temperature dependence of the magnetoelectric properties of rare-earth manganites \cite{Dong, 10.10631.3665634}, electric polarization appears below the characteristic temperature $T_H$. Upon further cooling, the magnetic moments of the rare-earth ions can undergo a separate long-range magnetic ordering, which gives rise to a small anomaly in the temperature dependence of the electric polarization. The microscopic origin of this anomaly has not yet been fully established. For heavy rare-earth ions, strong spin-orbit coupling and the resulting anisotropic g-factor tensor may significantly affect the electric polarization below the magnetic-ordering temperature of the rare-earth sublattice. Therefore, this mechanism may contribute to the experimentally observed ferroelectric anomaly in this temperature range.  

\subsection{Prediction of electric polarization for the helicoidal spin structure}
\subsubsection{The isotropic case}
Consider the helicoidal spin-density-wave state  
\begin{equation}\label{spin helix2}
    \bm{S}=S_{oy} cos(qx)\bm{e}_y+S_{oz} sin(qx)\bm{e}_z,
\end{equation}localized in the $yz$ plane and propagates perpendicular to this plane with the wave vector $\bm{q}=q\bm{e_x}$ (as we can see on the Fig. \eqref{fig2}). 
From the dependence of electric polarization on the spin density \eqref{IsotropicP}, for the isotropic scalar gyromagnetic ratio, the macroscopic electric polarization is equal to zero $\bm{P}=0$. 

\subsubsection{The tensor g-factor}
\begin{figure}
\includegraphics[width=8.4cm,angle=0]{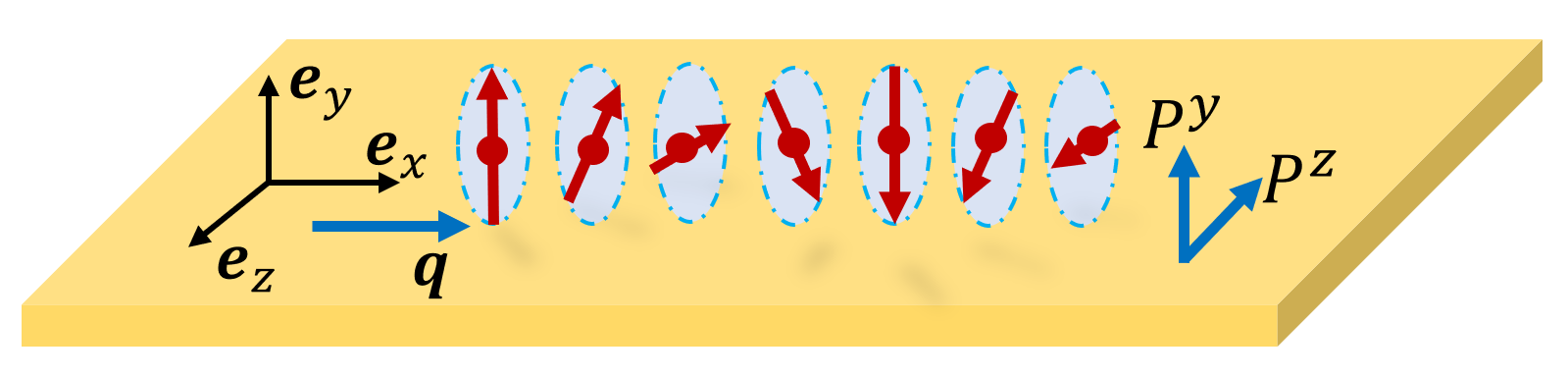}
\caption{\label{fig2} (Color online)
The helicoidal spiral spin structure in the absence of an external magnetic field \eqref{spin helix2}. The spin-density-wave state lies in the $yz$-plane with the wave vector $\bm{q}=(q, 0, 0)$ perpendicular to this plane. Two projections of electric polarization \eqref{projection of Pyx} and \eqref{projection of Pzx} appear  according to the definition \eqref{PHH}, and are determined by the non-diagonal components of the gyromagnetic ratio (g-factor).} 
\end{figure}
Substituting the helical spin structure \eqref{spin helix2} into the expression for the macroscopic electric polarization \eqref{PHH}, we obtain the polarization components lying in the spiral plane and perpendicular to the wave vector (see Fig. \eqref{fig2})
\begin{align}\label{projection of Pyx}
 P^y=-\frac{\gamma^{zx}}{c}\lambda_u\biggl(S^y\partial_xS^z - S^z\partial_x S^y\biggr)\qquad\nonumber\\ = -\gamma^{zx}\frac{\lambda_u}{c}qS_{oy}S_{oz},
\end{align}
\begin{align}\label{projection of Pzx}
 P^z=\frac{\gamma^{yx}}{c}\lambda_u\biggl(S^y\partial_xS^z - S^z\partial_x S^y\biggr)\qquad\nonumber\\ = \gamma^{yx}\frac{\lambda_u}{c}qS_{oy}S_{oz},   \\
 P^x=0.\qquad\qquad\qquad\qquad
\end{align}
According to Kimura, the first-order magnetoelectric effect is absent in screw-spiral magnets \cite{Kimura}. On the other hand, the electric polarization given by Eqs. \eqref{projection of Pyx} and \eqref{projection of Pzx}, obtained within the generalized spin-current model, is proportional to the off-diagonal components of the gyromagnetic-ratio tensor, $\gamma^{zx}$ and $\gamma^{yx}$. For a diagonal g-factor tensor, the electric polarization vanishes, in agreement with the phenomenological theory \cite{Dong}.

The polarization predicted here for helicoidal (proper-screw) spin structures originates from the anisotropy of the local g-tensor and therefore represents a mechanism distinct from the conventional inverse Dzyaloshinskii–Moriya effect \cite{Kimura2006}. In contrast to the latter, which vanishes for an ideal proper screw, the present mechanism becomes active when the local magnetic environment gives rise to off-diagonal components of the g-tensor. Such anisotropy naturally occurs in systems with strong spin-orbit coupling and low site symmetry. Candidate materials include proper-screw multiferroics \cite{Johnson2012} such as CuFeO$_2$, AgCrO$_2$, and RFe$_3$(BO$_3$)$_4$, where magnetic order develops in noncentrosymmetric or magnetically symmetry-broken phases and the magnetic ions occupy crystallographic sites of sufficiently low local symmetry. In these systems, anisotropic g-tensors have been reported or are expected owing to spin-orbit coupling, particularly for rare-earth ions or distorted transition-metal coordination polyhedra. In crystals containing several magnetic sublattices, the macroscopic polarization is obtained by summing the contributions from all sublattices. Depending on the symmetry relations between the local g-tensors, these contributions may either reinforce or compensate each other. Therefore, a quantitative prediction for a particular material requires a magnetic space-group analysis and knowledge of the local g-tensors, which is beyond the scope of the present general theory.

\section{Electric polarization caused by the Dzyaloshinskii–Moriya interaction}\label{Sec4}
\subsection{The cycloidal and helicoidal spin structures}
\subsubsection{The isotropic case}
For the isotropic g-factor and for the spin current, which is caused by the Dzyaloshinskii–Moriya interaction \cite{Andreev_2024, JETP} the expression of electric polarization  can be derived in the form
\begin{align}\label{PDMiso}
    P^\alpha=\frac{\gamma}{c}\varepsilon^{\alpha\mu\nu} J_{DM}^{\mu\nu}\qquad\qquad \nonumber\\ = \frac{\lambda_{\beta}}{c}\gamma \delta^\alpha \bm{S}^2,
\end{align} and is proportional to the vector of displacement of diamagnetic ions. We consider a sublattice of oxygen ions shifting in the same direction on the vector $\delta^\alpha$.  

Let us consider the cycloidal spin structure $\bm{S}=S_{ox} cos(qx)\bm{e}_x+S_{oy} sin(qx)\bm{e}_y$ and the helicoidal spin structure $\bm{S}=S_{oy} cos(qx)\bm{e}_y+S_{oz} sin(qx)\bm{e}_z$, with the wave vectors directed along the $x$ axis, while the oxygen ions are displaced along the $y$ direction, $\bm{\delta}=\delta \bm{e}_y$, i.e., within the plane of the spin spirals. The expression for the electric polarization \eqref{PDMiso} depends on the square of the spin-density vector and is determined by the direction of the displacement of the diamagnetic oxygen ions. Since the oxygen-ion displacement is chosen along the $y$ axis, the electric polarization for the scalar g-factor is also directed along the $y$ axis. We also take into account that the gyromagnetic ratio of the electron system is negative; therefore, the direction of the electric polarization is opposite to the displacement of the negatively charged oxygen ions. For the cycloidal and helicoidal spin structures, the macroscopic electric polarization takes the form
\begin{align} \label{PDMisoy}
    P^y=\frac{\gamma}{c}\lambda_{\beta}\delta S^2_o, \quad S^2_o=\sum_{i}S^2_{oi},
\end{align} and $i=x, y$ for the cycloidal, $i=y, z$ for the helicoidal spin orders.  

\subsubsection{The tensor g-factor}
 
If we consider the cycloidal \eqref{spin helix} and  helicoidal \eqref{spin helix2} spin structures with the wave vectors directed along the $x$ - axis, the projections of electric polarizations can be derived in the form
 \begin{align}\label{PDMx}
     P^x=-\frac{\lambda_{\beta}}{2c}\gamma^{yx}\cdot\delta\sum_{i}S_{oi}^2, \qquad\qquad\qquad\end{align} \begin{align}\label{PDMz} P^z=-\frac{\lambda_{\beta}}{2c}\gamma^{yz}\cdot\delta\sum_{i}S_{oi}^2, \qquad\end{align}  
   \begin{align}\label{PDMy}  \qquad\qquad P^y =\frac{\lambda_{\beta}}{2c}(\gamma^{xx}+\gamma^{zz})\cdot\delta\sum_{i}S_{oi}^2,    
 \end{align}where $i=x, y$ corresponds to the cycloidal spin order, and $i=y, z$ corresponds to the helicoidal spin order. The projection of the polarization onto the $y$-axis in Eq.~\eqref{PDMy} coincides with the isotropic case if we assume that $2\gamma=\gamma^{xx}+\gamma^{zz}$. However, the other polarization components in Eqs.~\eqref{PDMx} and \eqref{PDMz} are determined by the off-diagonal components of the gyromagnetic tensor and are absent in the isotropic case. This additional contribution may lead to a finite polarization when considering a sublattice of heavy magnetic ions for which the LS coupling no longer applies. 

 \section{The electric polarization caused by the Keffer-like form of the symmetric Heisenberg exchange interaction}\label{Sec5}
\subsection{The isotropic case}
For both symmetric exchange interactions and antisymmetric Dzyaloshinskii–Moriya interactions, the spin-current model has been used to describe the induced polarization in ferromagnetic spin structures. The symmetric Heisenberg exchange considered here does not require displacement of the diamagnetic oxygen ion, whereas the antisymmetric Dzyaloshinskii–Moriya interaction is realized solely through displacement of the nonmagnetic oxygen ion relative to the bond connecting two magnetic ions in the cluster. The obtained results can also be generalized to antiferromagnetic spin order. In this section, we consider the symmetric indirect exchange interaction mediated by an intermediate diamagnetic oxygen ion. This type of exchange coupling also leads to the formation of electric polarization in antiferromagnetic materials \cite{andreev2025keffer}.

A novel form of spin–spin interaction associated with the odd anisotropy of the symmetric exchange between magnetic ions mediated by a nonmagnetic ion is proposed in Ref.~\cite{andreev2025keffer}. The Heisenberg Hamiltonian of the indirect exchange interaction with a scalar coefficient can be written in the form
\begin{align}
\hat{H}=-\frac{1}{2}\sum_{i,j}U_{ij}\biggl(\hat{\bm{S}}_i\cdot\hat{\bm{S}}_j\biggr),
\end{align}
where the scalar coefficient, i.e., the exchange integral for the antiferromagnetic configuration, has the following expression
\begin{align}\label{U}
U_{ij}=U_{o,ij}+U_{1,ij} \biggl(\bm{r}_{ij}\cdot\bm{\delta}_{3,ij-AB}\biggr) \qquad\qquad\nonumber\\
+U_{2,ij}\biggl(\bm{r}_{ij}\cdot\biggl[\bm{\delta}_{2,ij-AB}\times \bm{\delta}_1\biggr]\biggr).
\end{align}
As can be seen, the exchange integral in Eq.~\eqref{U} has a complex structure, where $U_{o,ij}$ is the conventional exchange integral, while the second and third terms in the Hamiltonian represent the odd anisotropic part of the symmetric Heisenberg coupling and involve partial ligand displacements: $\bm{\delta}_1$, $\bm{\delta}_{2,ij-AB}$, which is perpendicular to the plane formed by the vectors $\bm{r}_{AB}$ and $\bm{\delta}_1$, and $\bm{\delta}_{3,ij-AB}$, which is parallel to the vector $\bm{r}_{AB}$ connecting the magnetic ions $A$ and $B$. The analysis of the noncollinear equilibrium state allows one to estimate the numerical values of $U_{1,ij}$ and $U_{2,ij}$ in terms of known parameters. The interaction under consideration can be written in a form that includes both mechanisms of ligand displacements as $U_{ij}=l_{ij}\delta^\beta_{eff,ij}r^\beta_{ij},$ where $\bm{r}_{ij}$ is the relative position vector between the magnetic ions, $l_{ij}=U_{1,ij}$ or $l_{ij}=U_{2,ij}$, and the effective displacement $\bm{\delta}_{eff}$ is introduced for each case as $\bm{\delta}_{eff}=\bm{\delta}_{3,ij-AB}$ or $\bm{\delta}_{eff}=\bm{\delta}_{2,ij-AB}\times\bm{\delta}_1$.

Following the spin-current formalism, the macroscopic polarization can be obtained from the generalized relation \eqref{PSC}
\begin{align}
P^{\mu}=\frac{\gamma}{c}\varepsilon^{\mu\alpha\beta}J^{\alpha\beta}_{\Sigma},
\end{align}
where the total spin current density $J^{\alpha\beta}_{\Sigma}=J^{\alpha\beta}_{A}+J^{\alpha\beta}_{B}$ is given by \cite{andreev2025keffer}
\begin{equation}\label{Jtotal}
J^{\alpha\beta}_{\Sigma} = \frac{1}{3}\varepsilon^{\alpha\mu\nu} g_{2l}\delta^\beta_{eff}S^\mu_A S^\nu_B,
\end{equation}
which finally yields the polarization
\begin{align}\label{Pkeffiso}
\bm{P} = \frac{1}{3}\frac{\gamma}{c} g_{2l} \biggl(\bm{S}_B\cdot(\bm{S}_A\cdot\bm{\delta}_{eff}) - \bm{S}_A\cdot(\bm{S}_B\cdot\bm{\delta}_{eff})\biggr),
\end{align}
where the effective ligand displacement is taken as $\bm{\delta}_{eff}=\bm{\delta}_{3,ij-AB}$ or $\bm{\delta}_{eff}=\bm{\delta}_{2,ij-AB}\times\bm{\delta}_1$, respectively, and $g_{2l}=\int l(\xi)\xi^2d\bm{\xi}$. The expression in the brackets describes the deviation of the spin projections along the effective ligand direction $\bm{\delta}_{eff}$ at fixed atomic positions.

\subsubsection{The cycloidal and helicoidal spin structures}
We consider noncollinear equilibrium spin state, which has the following spin densities of the magnetic sublattices $\bm{L}=\bm{S}_A-\bm{S}_B$ and $\bm{M}=\bm{S}_A+\bm{S}_B$ for the cycloidal state
\begin{align}\label{cycloida}
    \bm{L}=L_{ox}cos(qx)\bm{e}_x\pm L_{ox}sin(qx)\bm{e}_y,\qquad\qquad\\ \bm{M}=M_{ox}sin(qx)\bm{e}_x\mp M_{ox}cos(qx)\bm{e}_y,
\end{align}and for the helicoidal spin order
\begin{align}\label{helicoida}
   \bm{L}=L_{oy}cos(qx)\bm{e}_y\pm L_{oy}sin(qx)\bm{e}_z,\qquad\qquad\\ \bm{M}=M_{oy}sin(qx)\bm{e}_y\mp M_{oy}cos(qx)\bm{e}_z, 
\end{align} where $\bm{S}_A$ and $\bm{S}_B$ are the spin densities of the magnetic sublattices for the spin-up (A) and spin-down (B) states. The spin torque $T^\alpha \sim g_{2l}(\bm{\delta}_{eff}\cdot\bm{q}) f^\alpha(S_A, S_B)$ has nonzero projection on the wave vector of equilibrium
spin state when $\bm{\delta}_{eff}=\delta_{eff}\bm{e}_x$. For this shift of diamagnetic oxygen ions we obtain from the expression \eqref{Pkeffiso} the following formula for the electric polarization \begin{align}
    \bm{P}=\frac{\gamma g_{2l}\delta_{eff} }{6c} \biggl(\bm{M}\cdot L_x - \bm{L}\cdot M_x\biggr),
\end{align}
and only one projection of electric polarization for the cycloidal structure exists 
\begin{align}\label{KefferisoPy}
    P^y=\mp \frac{\gamma g_{2l}\delta_{eff}}{6c}L_{ox}M_{ox}, \qquad P^x=0, \qquad P^z=0. 
\end{align} In the isotropic case, for the helicoidal spin structures, macroscopic electric polarization is not induced $\bm{P}=0$.

\subsection{The case of tensor g-factor}
Let us consider the dependence of the electric polarization on the spin current density \eqref{PSC} for the tensor gyromagnetic ratio. Using the definition of the total spin current density \eqref{Jtotal}, we obtain the macroscopic polarization
\begin{equation}\label{Pkeff}
    P^\alpha=\frac{\gamma^{\mu\nu}}{3c}g_{2l}\varepsilon^{\alpha\mu\eta}\varepsilon^{\nu\beta\gamma}\delta^\eta_{eff}S^\beta_AS^\gamma_B.
\end{equation}
The cycloidal spin order leads to the emergence of the macroscopic electric polarization which is perpendicular to the direction of the wave vector and shift of oxygen ions
\begin{align}\label{KefferPy}
    P^y=\frac{g_{2l}\delta_{eff}}{6c}\gamma^{zz}\biggl(L_x\cdot M_y - L_y\cdot M_x\biggr)\qquad\nonumber\\=\mp \frac{\gamma^{zz} g_{2l}\delta_{eff}}{6c}L_{ox}M_{ox},
\end{align}
\begin{align}\label{KefferPz}
    P^z=-\frac{g_{2l}\delta_{eff}}{6c}\gamma^{yz}\biggl(L_x\cdot M_y - L_y\cdot M_x\biggr)\qquad\nonumber\\=\pm \frac{\gamma^{yz} g_{2l}\delta_{eff}}{6c}L_{ox}M_{ox},\\ P^x=0.\qquad\qquad\qquad\qquad
\end{align}The solution in Eq.~\eqref{KefferPy} coincides with the isotropic regime if $\gamma^{zz}=\gamma$. A new component of the electric polarization, which is absent in the isotropic case, is determined by the off-diagonal component of the g-factor tensor $\gamma^{yz}$ and is perpendicular to the plane of the cycloidal spin spiral in Eq.~\eqref{KefferPz}.

Unlike the isotropic case, electric polarization can also be induced in the helicoidal spin order, and its vector lies in the plane of the spin spiral:
\begin{align}
P^y=\mp \frac{g_{2l}\delta_{eff}}{6c}\gamma^{zx} L_{oy}M_{oy},\qquad\\
P^z=\pm \frac{g_{2l}\delta_{eff}}{6c}\gamma^{yx} L_{oy}M_{oy}, \qquad
P^x=0.
\end{align}
These results demonstrate that the ligand displacement, which is usually responsible for the emergence of the Dzyaloshinskii–Moriya interaction, also affects the symmetric Heisenberg exchange in antiferromagnetic materials, leading to the emergence of a magnetoelectric effect of purely spin origin.

\section{Discussion}

The physical interpretation of the proposed mechanism can be understood within the framework of many-particle quantum hydrodynamics with a g-factor tensor. The symmetric Heisenberg exchange interaction does not directly produce electric polarization. In general, it determines the equilibrium spin texture, while the corresponding effective spin-current tensor naturally emerges from the microscopic Hamiltonian as a macroscopic hydrodynamic quantity \cite{Andreev_2024, Andreev_2019}. In contrast, the antisymmetric Dzyaloshinskii-Moriya interaction favors noncollinear magnetic configurations owing to spin-orbit coupling in the absence of local inversion symmetry and often stabilizes cycloidal or helical spin structures \cite{Sergienko2006}. Within the present formalism, the spin-orbit interaction enters the momentum-balance equation through a force density proportional to the effective spin-current tensor. As shown, equilibrium between this relativistic force and the electrostatic force acting on the electric dipoles leads to the constitutive relation between the electric polarization and the effective spin current. Therefore, the emergence of electric polarization requires two essential conditions: a nonvanishing effective spin-current distribution established by exchange magnetic interactions and relativistic spin-orbit coupling that converts this spin current into macroscopic electric polarization.

The symmetry of the crystal further determines which components of the induced polarization are allowed. In particular, inversion symmetry must be broken either locally or by the magnetic order, whereas the tensor structure of the g-factor is governed by the local crystal-field symmetry of the magnetic ion \cite{Rubins2014}. Although any symmetric g-factor tensor can always be diagonalized in its local principal-axis system, these principal axes generally do not coincide with the crystallographic axes because they are determined by the local ligand environment. Since macroscopic electric polarization is measured in the crystallographic reference frame, the off-diagonal components of the g-factor tensor expressed in this basis remain physically observable and cannot, in general, be eliminated by a global coordinate transformation. Such a situation is common in low-symmetry multiferroics, where the local point symmetry of the magnetic ion is lower than the crystal symmetry.

These considerations define the range of applicability of the present model. In addition to the existence of spin-driven ferroelectricity, the model requires an anisotropic tensor g-factor originating from spin-orbit coupling and crystal-field effects. Such conditions are naturally realized in multiferroics containing transition-metal or rare-earth ions, where orbital contributions to the magnetic moment remain significant. In particular, rare-earth manganites exhibit strongly anisotropic magnetic properties associated with the $4f$ electrons of Tb$^{3+}$ and Dy$^{3+}$ ions, making a tensorial description of the g-factor essential for interpreting magnetic ordering and magnetic excitations \cite{PhysRevB.90.224418,LEE2021147013}. Although the average crystal symmetry of orthorhombic compounds often allows a diagonal $g$-tensor, the local symmetry of rare-earth ions is generally lower, resulting in finite off-diagonal components in the crystallographic coordinate system. Thus, the emergence of an anisotropic tensor g-factor requires several conditions. First, the magnetic ion must possess spin-orbit coupling, which couples the spin and orbital degrees of freedom. Second, the local crystal field must be anisotropic, lifting the degeneracy of the electronic states and defining preferred magnetic axes. Third, the local symmetry of the magnetic-ion site should be sufficiently low, or the principal axes of the local crystal field should be rotated with respect to the crystallographic axes, so that off-diagonal components of the g-tensor become finite in the crystallographic coordinate system.

A representative example is MnWO$_4$, a type-II multiferroic whose ferroelectric AF2 phase is characterized by an incommensurate noncollinear spin structure \cite{Heyer_2006}. Because the local principal axes of the magnetic ions do not coincide with the crystallographic axes, the off-diagonal components of the g-factor tensor can become experimentally observable in electron paramagnetic resonance, antiferromagnetic resonance, and magnetic-field-dependent measurements \cite{LI2026173724}. Similar considerations apply to LiCuVO$_4$, where the inverse Dzyaloshinskii–Moriya mechanism successfully explains the existence of spin-induced ferroelectricity, while quantitative agreement with experiment requires additional lattice and electronic contributions beyond the conventional KNB model.

Likewise, the original KNB model predicts that an ideal proper-screw spiral should not generate macroscopic electric polarization \cite{PhysRevLett.95.057205}. The generalized spin-current model developed in the present work is expected to be relevant for multiferroic materials that combine a helicoidal (proper-screw) magnetic structure with pronounced anisotropy of the magnetic response \cite{Arima2007}. Such conditions are most likely to be realized in compounds containing transition-metal or rare-earth ions with strong spin-orbit coupling, where the magnetic moments are characterized by a tensor g-factor determined by the local crystal-field environment. Candidate systems include proper-screw multiferroics such as CuFeO$_2$, AgCrO$_2$, CuCrO$_2$, and rare-earth ferroborates $R$Fe$_3$(BO$_3$)$_4$ \cite{Johnson2012}, in which noncollinear magnetic order develops in crystallographic environments of reduced local symmetry. In these materials, the principal axes of the local g-factor tensor generally do not coincide with the crystallographic axes, leading to finite off-diagonal components of the g-tensor in the crystallographic coordinate system. Since the conventional inverse Dzyaloshinskii-Moriya mechanism predicts no macroscopic polarization for an ideal proper-screw spiral, these compounds provide promising platforms for testing additional symmetry-allowed contributions to the electric polarization arising from the tensor nature of the g-factor.

\section{Conclusion}\label{Sec6}

The present work introduces, for the first time, a tensor g-factor into the microscopic many-particle Hamiltonian and develops the corresponding many-particle quantum-hydrodynamic description. This extension predicts qualitatively new contributions to the electric polarization originating from the off-diagonal components of the g-factor tensor and demonstrates how the local crystal-field symmetry of magnetic ions modifies the spin-current mechanism.

The proposed approach differs fundamentally from phenomenological Ginzburg-Landau theories. Within the phenomenological framework, the magnetoelectric coupling is constructed solely on the basis of symmetry arguments and is expressed in terms of macroscopic order parameters with phenomenological coupling coefficients. In contrast, our theory is derived directly from the microscopic many-particle Hamiltonian within the framework of many-particle quantum hydrodynamics. The tensor g-factor is incorporated at the Hamiltonian level, and the electric polarization follows from the momentum- and spin-balance equations without introducing phenomenological free-energy expansions.

As a consequence, the anisotropy of the local g-tensor enters explicitly into the microscopic expression for the effective spin current and gives rise to additional components of the electric polarization. Their microscopic origin can be traced directly to the combined effects of the local crystal field and spin-orbit coupling acting on the magnetic ions. In this sense, the tensor g-factor is not an additional phenomenological parameter but an intrinsic microscopic characteristic of the magnetic ions that modifies the effective spin current and, consequently, the spin-induced electric polarization.

In this paper, a mechanism of the magnetoelectric effect based on effective spin currents induced by different types of interactions has been theoretically developed. We have formulated and justified the spin-current model of Katsura, Nagaosa, and Balatsky (KNB) by incorporating a tensor form of the gyromagnetic ratio (or g-factor) using the basic principles of the many-particle quantum hydrodynamics method. Our description is based on a cluster model consisting of two positively charged magnetic ions with a diamagnetic, negatively charged oxygen ion $O^{2-}$ located between them, where each cluster is treated as a point-like quasiparticle within the many-particle quantum hydrodynamics framework. However, in contrast to the KNB approach, the oxygen ion is allowed to shift away from the midpoint of the bond between the magnetic ions. Using the results of the investigated model, we have analyzed the main mechanisms of spin-induced ferroelectricity.

The first mechanism is the symmetric Heisenberg exchange interaction. In the spin current induced by this type of interaction, the displacement of oxygen ions is not explicitly required for the emergence of polarization, although it is essential for the antisymmetric Dzyaloshinskii–Moriya interaction, which leads to oxygen-ion displacement and the formation of a canted spin structure. We demonstrate that the symmetric exchange coupling gives rise to an effective spin current, which in turn produces electric polarization in cycloidal and screw (helicoidal) spin systems. However, unlike the isotropic case in Eq.~\eqref{IsotropicP}, where cycloidal spin order and a scalar g-factor yield only a single component of the electric polarization in Eq.~\eqref{IsotropicPy}, the nondiagonal components of the g-factor tensor give rise to additional out-of-plane polarization components in Eqs.~\eqref{projection of Py} and \eqref{projection of Pz}. For the screw spin order, we obtain nonzero components of the electric polarization in Eqs.~\eqref{projection of Pyx} and \eqref{projection of Pzx}, also originating from the off-diagonal elements of the gyromagnetic-ratio tensor, which is atypical for the isotropic case.

In the second mechanism, the electric polarization induced by the Dzyaloshinskii–Moriya interaction is determined by the oxygen displacement vector from the midpoint of the bond and by the square of the spin density. This implies that such polarization can arise even in systems with collinear spins, in contrast to the KNB spin-current model, where only noncollinear spin configurations are considered and the polarization originates from the deformation of the electronic density. Solutions have also been derived for cycloidal and helicoidal spin orders in Eqs.~\eqref{PDMx}–\eqref{PDMy}. We have demonstrated the results of the model for the ferromagnetic state, and the model can be extended to antiferromagnetic orders.

In addition, we have considered a new mechanism for the formation of electric polarization caused by a novel form of spin–spin interaction related to the odd anisotropy of the symmetric exchange. This indirect exchange is mediated by the oxygen ion, which is displaced from the midpoint of the bond. We have derived the electric polarization for the isotropic scalar g-factor in Eq.~\eqref{Pkeffiso} and generalized the result to the case of a tensor gyromagnetic ratio in Eq.~\eqref{Pkeff}. New solutions for the electric polarization induced by this type of interaction have been predicted in Eqs.~\eqref{KefferisoPy}, \eqref{KefferPy}, and \eqref{KefferPz}.

\emph{DATA AVAILABILITY}:
Data sharing is not applicable to this article as no new data were created or analyzed in this study, which is a purely theoretical one.


\emph{Acknowledgements}:
The research is supported by the Russian Science Foundation under the grant No. 25-22-00064.
 
\nocite{_}


\begin{thebibliography}{59}%
\makeatletter
\providecommand \@ifxundefined [1]{%
 \@ifx{#1\undefined}
}%
\providecommand \@ifnum [1]{%
 \ifnum #1\expandafter \@firstoftwo
 \else \expandafter \@secondoftwo
 \fi
}%
\providecommand \@ifx [1]{%
 \ifx #1\expandafter \@firstoftwo
 \else \expandafter \@secondoftwo
 \fi
}%
\providecommand \natexlab [1]{#1}%
\providecommand \enquote  [1]{``#1''}%
\providecommand \bibnamefont  [1]{#1}%
\providecommand \bibfnamefont [1]{#1}%
\providecommand \citenamefont [1]{#1}%
\providecommand \href@noop [0]{\@secondoftwo}%
\providecommand \href [0]{\begingroup \@sanitize@url \@href}%
\providecommand \@href[1]{\@@startlink{#1}\@@href}%
\providecommand \@@href[1]{\endgroup#1\@@endlink}%
\providecommand \@sanitize@url [0]{\catcode `\\12\catcode `\$12\catcode `\&12\catcode `\#12\catcode `\^12\catcode `\_12\catcode `\%12\relax}%
\providecommand \@@startlink[1]{}%
\providecommand \@@endlink[0]{}%
\providecommand \url  [0]{\begingroup\@sanitize@url \@url }%
\providecommand \@url [1]{\endgroup\@href {#1}{\urlprefix }}%
\providecommand \urlprefix  [0]{URL }%
\providecommand \Eprint [0]{\href }%
\providecommand \doibase [0]{https://doi.org/}%
\providecommand \selectlanguage [0]{\@gobble}%
\providecommand \bibinfo  [0]{\@secondoftwo}%
\providecommand \bibfield  [0]{\@secondoftwo}%
\providecommand \translation [1]{[#1]}%
\providecommand \BibitemOpen [0]{}%
\providecommand \bibitemStop [0]{}%
\providecommand \bibitemNoStop [0]{.\EOS\space}%
\providecommand \EOS [0]{\spacefactor3000\relax}%
\providecommand \BibitemShut  [1]{\csname bibitem#1\endcsname}%
\let\auto@bib@innerbib\@empty
\bibitem [{\citenamefont {Kopyl}\ \emph {et~al.}(2021)\citenamefont {Kopyl}, \citenamefont {Surmenev}, \citenamefont {Surmeneva}, \citenamefont {Fetisov},\ and\ \citenamefont {Kholkin}}]{KOPYL2021100149}%
  \BibitemOpen
  \bibfield  {author} {\bibinfo {author} {\bibfnamefont {S.}~\bibnamefont {Kopyl}}, \bibinfo {author} {\bibfnamefont {R.}~\bibnamefont {Surmenev}}, \bibinfo {author} {\bibfnamefont {M.}~\bibnamefont {Surmeneva}}, \bibinfo {author} {\bibfnamefont {Y.}~\bibnamefont {Fetisov}},\ and\ \bibinfo {author} {\bibfnamefont {A.}~\bibnamefont {Kholkin}},\ }\bibfield  {title} {\bibinfo {title} {Magnetoelectric effect: principles and applications in biology and medicine– a review},\ }\href {https://doi.org/https://doi.org/10.1016/j.mtbio.2021.100149} {\bibfield  {journal} {\bibinfo  {journal} {Materials Today Bio}\ }\textbf {\bibinfo {volume} {12}},\ \bibinfo {pages} {100149} (\bibinfo {year} {2021})}\BibitemShut {NoStop}%
\bibitem [{\citenamefont {Khomskii}(2021)}]{Khomski}%
  \BibitemOpen
  \bibfield  {author} {\bibinfo {author} {\bibfnamefont {D.~I.}\ \bibnamefont {Khomskii}},\ }\bibfield  {title} {\bibinfo {title} {Multiferroics and beyond: Electric properties of different magnetic textures},\ }\href {https://doi.org/https://doi.org/10.1134/S1063776121040099} {\bibfield  {journal} {\bibinfo  {journal} {JETP}\ }\textbf {\bibinfo {volume} {132}},\ \bibinfo {pages} {482} (\bibinfo {year} {2021})}\BibitemShut {NoStop}%
\bibitem [{\citenamefont {Dong}\ and\ \citenamefont {et~al.}(2019)}]{Dong}%
  \BibitemOpen
  \bibfield  {author} {\bibinfo {author} {\bibfnamefont {S.}~\bibnamefont {Dong}}\ and\ \bibinfo {author} {\bibnamefont {et~al.}},\ }\bibfield  {title} {\bibinfo {title} {Magnetoelectricity in multiferroics: a theoretical perspective},\ }\href {https://doi.org/https://doi.org/10.1093/nsr/nwz023} {\bibfield  {journal} {\bibinfo  {journal} {Natl. Sci. Rev.}\ }\textbf {\bibinfo {volume} {6}},\ \bibinfo {pages} {629} (\bibinfo {year} {2019})}\BibitemShut {NoStop}%
\bibitem [{\citenamefont {Andreev}\ and\ \citenamefont {Trukhanova}(2024{\natexlab{a}})}]{Andreev}%
  \BibitemOpen
  \bibfield  {author} {\bibinfo {author} {\bibfnamefont {P.}~\bibnamefont {Andreev}}\ and\ \bibinfo {author} {\bibfnamefont {M.~I.}\ \bibnamefont {Trukhanova}},\ }\bibfield  {title} {\bibinfo {title} {Polarization evolution equation for exchange-strictionally formed type ii multiferroic materials},\ }\href {https://doi.org/https://doi.org/10.1140/epjb/s10051-024-00756-7} {\bibfield  {journal} {\bibinfo  {journal} {Eur. Phys. J. B}\ }\textbf {\bibinfo {volume} {97}},\ \bibinfo {pages} {116} (\bibinfo {year} {2024}{\natexlab{a}})}\BibitemShut {NoStop}%
\bibitem [{\citenamefont {Khomskii}(2009)}]{Khomski2}%
  \BibitemOpen
  \bibfield  {author} {\bibinfo {author} {\bibfnamefont {D.~I.}\ \bibnamefont {Khomskii}},\ }\bibfield  {title} {\bibinfo {title} {Classifying multiferroics: Mechanisms and effects},\ }\href {https://physics.aps.org/articles/v2/20} {\bibfield  {journal} {\bibinfo  {journal} {Physics}\ }\textbf {\bibinfo {volume} {2}},\ \bibinfo {pages} {20} (\bibinfo {year} {2009})}\BibitemShut {NoStop}%
\bibitem [{\citenamefont {Neaton}\ \emph {et~al.}(2005)\citenamefont {Neaton}, \citenamefont {Ederer}, \citenamefont {Waghmare}, \citenamefont {Spaldin},\ and\ \citenamefont {Rabe}}]{PhysRevB.71.014113}%
  \BibitemOpen
  \bibfield  {author} {\bibinfo {author} {\bibfnamefont {J.~B.}\ \bibnamefont {Neaton}}, \bibinfo {author} {\bibfnamefont {C.}~\bibnamefont {Ederer}}, \bibinfo {author} {\bibfnamefont {U.~V.}\ \bibnamefont {Waghmare}}, \bibinfo {author} {\bibfnamefont {N.~A.}\ \bibnamefont {Spaldin}},\ and\ \bibinfo {author} {\bibfnamefont {K.~M.}\ \bibnamefont {Rabe}},\ }\bibfield  {title} {\bibinfo {title} {First-principles study of spontaneous polarization in multiferroic $\mathrm{Bi}\mathrm{Fe}\mathrm{O}_{3}$},\ }\href {https://doi.org/10.1103/PhysRevB.71.014113} {\bibfield  {journal} {\bibinfo  {journal} {Phys. Rev. B}\ }\textbf {\bibinfo {volume} {71}},\ \bibinfo {pages} {014113} (\bibinfo {year} {2005})}\BibitemShut {NoStop}%
\bibitem [{\citenamefont {Ederer}\ and\ \citenamefont {Spaldin}(2005)}]{PhysRevB.71.060401}%
  \BibitemOpen
  \bibfield  {author} {\bibinfo {author} {\bibfnamefont {C.}~\bibnamefont {Ederer}}\ and\ \bibinfo {author} {\bibfnamefont {N.~A.}\ \bibnamefont {Spaldin}},\ }\bibfield  {title} {\bibinfo {title} {Weak ferromagnetism and magnetoelectric coupling in bismuth ferrite},\ }\href {https://doi.org/10.1103/PhysRevB.71.060401} {\bibfield  {journal} {\bibinfo  {journal} {Phys. Rev. B}\ }\textbf {\bibinfo {volume} {71}},\ \bibinfo {pages} {060401} (\bibinfo {year} {2005})}\BibitemShut {NoStop}%
\bibitem [{\citenamefont {Catalan}\ and\ \citenamefont {Scott}(2009)}]{Catalan}%
  \BibitemOpen
  \bibfield  {author} {\bibinfo {author} {\bibfnamefont {G.}~\bibnamefont {Catalan}}\ and\ \bibinfo {author} {\bibfnamefont {J.~F.}\ \bibnamefont {Scott}},\ }\bibfield  {title} {\bibinfo {title} {Physics and applications of bismuth ferrite},\ }\href {https://doi.org/https://doi.org/10.1002/adma.200802849} {\bibfield  {journal} {\bibinfo  {journal} {Advanced Materials}\ }\textbf {\bibinfo {volume} {21}},\ \bibinfo {pages} {2463} (\bibinfo {year} {2009})}\BibitemShut {NoStop}%
\bibitem [{\citenamefont {Kimura}(2007)}]{Kimura}%
  \BibitemOpen
  \bibfield  {author} {\bibinfo {author} {\bibfnamefont {T.}~\bibnamefont {Kimura}},\ }\bibfield  {title} {\bibinfo {title} {Spiral magnets as magnetoelectrics},\ }\href {https://doi.org/10.1146/annurev.matsci.37.052506.084259} {\bibfield  {journal} {\bibinfo  {journal} {Annual Review of Materials Research}\ }\textbf {\bibinfo {volume} {37}},\ \bibinfo {pages} {387} (\bibinfo {year} {2007})}\BibitemShut {NoStop}%
\bibitem [{\citenamefont {Kimura}\ \emph {et~al.}(2005)\citenamefont {Kimura}, \citenamefont {Lawes}, \citenamefont {Goto}, \citenamefont {Tokura},\ and\ \citenamefont {Ramirez}}]{Kimura2}%
  \BibitemOpen
  \bibfield  {author} {\bibinfo {author} {\bibfnamefont {T.}~\bibnamefont {Kimura}}, \bibinfo {author} {\bibfnamefont {G.~J.}\ \bibnamefont {Lawes}}, \bibinfo {author} {\bibfnamefont {T.}~\bibnamefont {Goto}}, \bibinfo {author} {\bibfnamefont {Y.}~\bibnamefont {Tokura}},\ and\ \bibinfo {author} {\bibfnamefont {A.~P.}\ \bibnamefont {Ramirez}},\ }\bibfield  {title} {\bibinfo {title} {Magnetoelectric phase diagrams of orthorhombic {RMnO$_3$} ({R} = {Gd} , {Tb}, and {Dy})},\ }\href {https://link.aps.org/doi/10.1103/PhysRevB.71.224425} {\bibfield  {journal} {\bibinfo  {journal} {Phys. Rev. B}\ }\textbf {\bibinfo {volume} {71}},\ \bibinfo {pages} {224425} (\bibinfo {year} {2005})}\BibitemShut {NoStop}%
\bibitem [{\citenamefont {Mostovoy}(2006)}]{PhysRevLett.96.067601}%
  \BibitemOpen
  \bibfield  {author} {\bibinfo {author} {\bibfnamefont {M.}~\bibnamefont {Mostovoy}},\ }\bibfield  {title} {\bibinfo {title} {Ferroelectricity in spiral magnets},\ }\href {https://doi.org/10.1103/PhysRevLett.96.067601} {\bibfield  {journal} {\bibinfo  {journal} {Phys. Rev. Lett.}\ }\textbf {\bibinfo {volume} {96}},\ \bibinfo {pages} {067601} (\bibinfo {year} {2006})}\BibitemShut {NoStop}%
\bibitem [{\citenamefont {Pyatakov}\ and\ \citenamefont {Zvezdin}(2012)}]{Pyatakov}%
  \BibitemOpen
  \bibfield  {author} {\bibinfo {author} {\bibfnamefont {A.~P.}\ \bibnamefont {Pyatakov}}\ and\ \bibinfo {author} {\bibfnamefont {A.~K.}\ \bibnamefont {Zvezdin}},\ }\bibfield  {title} {\bibinfo {title} {Magnetoelectric and multiferroic media},\ }\href {https://doi.org/10.3367/UFNe. 0182.201206b.0593} {\bibfield  {journal} {\bibinfo  {journal} {Phys. Usp.}\ }\textbf {\bibinfo {volume} {55}},\ \bibinfo {pages} {557} (\bibinfo {year} {2012})}\BibitemShut {NoStop}%
\bibitem [{\citenamefont {Tokura}\ \emph {et~al.}(2014)\citenamefont {Tokura}, \citenamefont {Seki},\ and\ \citenamefont {Nagaosa}}]{Tokura}%
  \BibitemOpen
  \bibfield  {author} {\bibinfo {author} {\bibfnamefont {Y.}~\bibnamefont {Tokura}}, \bibinfo {author} {\bibfnamefont {S.}~\bibnamefont {Seki}},\ and\ \bibinfo {author} {\bibfnamefont {N.}~\bibnamefont {Nagaosa}},\ }\bibfield  {title} {\bibinfo {title} {Multiferroics of spin origin},\ }\href {https://doi.org/10.1088/0034-4885/77/7/076501} {\bibfield  {journal} {\bibinfo  {journal} {Rep. Prog. Phys.}\ }\textbf {\bibinfo {volume} {77}},\ \bibinfo {pages} {076501} (\bibinfo {year} {2014})}\BibitemShut {NoStop}%
\bibitem [{\citenamefont {Okuyama}\ \emph {et~al.}(2011)\citenamefont {Okuyama}, \citenamefont {Ishiwata}, \citenamefont {Takahashi}, \citenamefont {Yamauchi}, \citenamefont {Picozzi}, \citenamefont {Sugimoto}, \citenamefont {Sakai}, \citenamefont {Takata}, \citenamefont {Shimano}, \citenamefont {Taguchi}, \citenamefont {Arima},\ and\ \citenamefont {Tokura}}]{PhysRevB.84.054440}%
  \BibitemOpen
  \bibfield  {author} {\bibinfo {author} {\bibfnamefont {D.}~\bibnamefont {Okuyama}}, \bibinfo {author} {\bibfnamefont {S.}~\bibnamefont {Ishiwata}}, \bibinfo {author} {\bibfnamefont {Y.}~\bibnamefont {Takahashi}}, \bibinfo {author} {\bibfnamefont {K.}~\bibnamefont {Yamauchi}}, \bibinfo {author} {\bibfnamefont {S.}~\bibnamefont {Picozzi}}, \bibinfo {author} {\bibfnamefont {K.}~\bibnamefont {Sugimoto}}, \bibinfo {author} {\bibfnamefont {H.}~\bibnamefont {Sakai}}, \bibinfo {author} {\bibfnamefont {M.}~\bibnamefont {Takata}}, \bibinfo {author} {\bibfnamefont {R.}~\bibnamefont {Shimano}}, \bibinfo {author} {\bibfnamefont {Y.}~\bibnamefont {Taguchi}}, \bibinfo {author} {\bibfnamefont {T.}~\bibnamefont {Arima}},\ and\ \bibinfo {author} {\bibfnamefont {Y.}~\bibnamefont {Tokura}},\ }\bibfield  {title} {\bibinfo {title} {Magnetically driven ferroelectric atomic displacements in orthorhombic ymno${}_{3}$},\ }\href {https://doi.org/10.1103/PhysRevB.84.054440} {\bibfield  {journal} {\bibinfo  {journal} {Phys. Rev. B}\
  }\textbf {\bibinfo {volume} {84}},\ \bibinfo {pages} {054440} (\bibinfo {year} {2011})}\BibitemShut {NoStop}%
\bibitem [{\citenamefont {Wu}\ \emph {et~al.}(2009)\citenamefont {Wu}, \citenamefont {Burnus}, \citenamefont {Hu}, \citenamefont {Martin}, \citenamefont {Maignan}, \citenamefont {Cezar}, \citenamefont {Tanaka}, \citenamefont {Brookes}, \citenamefont {Khomskii},\ and\ \citenamefont {Tjeng}}]{PhysRevLett.102.026404}%
  \BibitemOpen
  \bibfield  {author} {\bibinfo {author} {\bibfnamefont {H.}~\bibnamefont {Wu}}, \bibinfo {author} {\bibfnamefont {T.}~\bibnamefont {Burnus}}, \bibinfo {author} {\bibfnamefont {Z.}~\bibnamefont {Hu}}, \bibinfo {author} {\bibfnamefont {C.}~\bibnamefont {Martin}}, \bibinfo {author} {\bibfnamefont {A.}~\bibnamefont {Maignan}}, \bibinfo {author} {\bibfnamefont {J.~C.}\ \bibnamefont {Cezar}}, \bibinfo {author} {\bibfnamefont {A.}~\bibnamefont {Tanaka}}, \bibinfo {author} {\bibfnamefont {N.~B.}\ \bibnamefont {Brookes}}, \bibinfo {author} {\bibfnamefont {D.~I.}\ \bibnamefont {Khomskii}},\ and\ \bibinfo {author} {\bibfnamefont {L.~H.}\ \bibnamefont {Tjeng}},\ }\bibfield  {title} {\bibinfo {title} {Ising magnetism and ferroelectricity in ${\mathrm{ca}}_{3}{\mathrm{comno}}_{6}$},\ }\href {https://doi.org/10.1103/PhysRevLett.102.026404} {\bibfield  {journal} {\bibinfo  {journal} {Phys. Rev. Lett.}\ }\textbf {\bibinfo {volume} {102}},\ \bibinfo {pages} {026404} (\bibinfo {year} {2009})}\BibitemShut {NoStop}%
\bibitem [{\citenamefont {Katsura}\ \emph {et~al.}(2005)\citenamefont {Katsura}, \citenamefont {Nagaosa},\ and\ \citenamefont {Balatsky}}]{PhysRevLett.95.057205}%
  \BibitemOpen
  \bibfield  {author} {\bibinfo {author} {\bibfnamefont {H.}~\bibnamefont {Katsura}}, \bibinfo {author} {\bibfnamefont {N.}~\bibnamefont {Nagaosa}},\ and\ \bibinfo {author} {\bibfnamefont {A.~V.}\ \bibnamefont {Balatsky}},\ }\bibfield  {title} {\bibinfo {title} {Spin current and magnetoelectric effect in noncollinear magnets},\ }\href {https://doi.org/10.1103/PhysRevLett.95.057205} {\bibfield  {journal} {\bibinfo  {journal} {Phys. Rev. Lett.}\ }\textbf {\bibinfo {volume} {95}},\ \bibinfo {pages} {057205} (\bibinfo {year} {2005})}\BibitemShut {NoStop}%
\bibitem [{\citenamefont {Sergienko}\ and\ \citenamefont {Dagotto}(2006{\natexlab{a}})}]{PhysRevB.73.094434}%
  \BibitemOpen
  \bibfield  {author} {\bibinfo {author} {\bibfnamefont {I.~A.}\ \bibnamefont {Sergienko}}\ and\ \bibinfo {author} {\bibfnamefont {E.}~\bibnamefont {Dagotto}},\ }\bibfield  {title} {\bibinfo {title} {Role of the dzyaloshinskii-moriya interaction in multiferroic perovskites},\ }\href {https://doi.org/10.1103/PhysRevB.73.094434} {\bibfield  {journal} {\bibinfo  {journal} {Phys. Rev. B}\ }\textbf {\bibinfo {volume} {73}},\ \bibinfo {pages} {094434} (\bibinfo {year} {2006}{\natexlab{a}})}\BibitemShut {NoStop}%
\bibitem [{\citenamefont {Trukhanova}\ \emph {et~al.}(2025)\citenamefont {Trukhanova}, \citenamefont {Andreev},\ and\ \citenamefont {Obukhov}}]{10.1142/S0217979225500729}%
  \BibitemOpen
  \bibfield  {author} {\bibinfo {author} {\bibfnamefont {M.~I.}\ \bibnamefont {Trukhanova}}, \bibinfo {author} {\bibfnamefont {P.}~\bibnamefont {Andreev}},\ and\ \bibinfo {author} {\bibfnamefont {Y.~N.}\ \bibnamefont {Obukhov}},\ }\bibfield  {title} {\bibinfo {title} {A new quantum hydrodynamic description of ferroelectricity in spiral magnets},\ }\href {https://doi.org/10.1142/S0217979225500729} {\bibfield  {journal} {\bibinfo  {journal} {Int. J. Mod. Phys. B}\ }\textbf {\bibinfo {volume} {39}},\ \bibinfo {pages} {2550072} (\bibinfo {year} {2025})}\BibitemShut {NoStop}%
\bibitem [{\citenamefont {Andreev}\ and\ \citenamefont {Trukhanova}(2024{\natexlab{b}})}]{Andreev_2024}%
  \BibitemOpen
  \bibfield  {author} {\bibinfo {author} {\bibfnamefont {P.~A.}\ \bibnamefont {Andreev}}\ and\ \bibinfo {author} {\bibfnamefont {M.~I.}\ \bibnamefont {Trukhanova}},\ }\bibfield  {title} {\bibinfo {title} {Electric polarization evolution equation for antiferromagnetic multiferroics with the polarization proportional to the scalar product of the spins},\ }\href {https://doi.org/10.1088/1402-4896/ad7a35} {\bibfield  {journal} {\bibinfo  {journal} {Physica Scripta}\ }\textbf {\bibinfo {volume} {99}},\ \bibinfo {pages} {1059b2} (\bibinfo {year} {2024}{\natexlab{b}})}\BibitemShut {NoStop}%
\bibitem [{\citenamefont {Andreev}\ and\ \citenamefont {Trukhanova}(2024{\natexlab{c}})}]{Trukhanova_2024}%
  \BibitemOpen
  \bibfield  {author} {\bibinfo {author} {\bibfnamefont {P.~A.}\ \bibnamefont {Andreev}}\ and\ \bibinfo {author} {\bibfnamefont {M.~I.}\ \bibnamefont {Trukhanova}},\ }\bibfield  {title} {\bibinfo {title} {Polarization evolution equation for exchange-strictionally formed type ii multiferroic materials},\ }\href {https://doi.org/10.1140/epjb/s10051-024-00756-7} {\bibfield  {journal} {\bibinfo  {journal} {Eur. Phys. J. B}\ }\textbf {\bibinfo {volume} {97}},\ \bibinfo {pages} {116} (\bibinfo {year} {2024}{\natexlab{c}})}\BibitemShut {NoStop}%
\bibitem [{\citenamefont {Schwan}\ \emph {et~al.}(2011)\citenamefont {Schwan}, \citenamefont {Meiners}, \citenamefont {Greilich}, \citenamefont {Yakovlev}, \citenamefont {Bayer}, \citenamefont {Maia}, \citenamefont {Quivy},\ and\ \citenamefont {Henriques}}]{10.1063/1.3665634}%
  \BibitemOpen
  \bibfield  {author} {\bibinfo {author} {\bibfnamefont {A.}~\bibnamefont {Schwan}}, \bibinfo {author} {\bibfnamefont {B.-M.}\ \bibnamefont {Meiners}}, \bibinfo {author} {\bibfnamefont {A.}~\bibnamefont {Greilich}}, \bibinfo {author} {\bibfnamefont {D.~R.}\ \bibnamefont {Yakovlev}}, \bibinfo {author} {\bibfnamefont {M.}~\bibnamefont {Bayer}}, \bibinfo {author} {\bibfnamefont {A.~D.~B.}\ \bibnamefont {Maia}}, \bibinfo {author} {\bibfnamefont {A.~A.}\ \bibnamefont {Quivy}},\ and\ \bibinfo {author} {\bibfnamefont {A.~B.}\ \bibnamefont {Henriques}},\ }\bibfield  {title} {\bibinfo {title} {Anisotropy of electron and hole g-factors in ($\mathrm{I}$n,$\mathrm{G}$a)$\mathrm{A}$s quantum dots},\ }\href {https://doi.org/10.1063/1.3665634} {\bibfield  {journal} {\bibinfo  {journal} {Appl. Phys. Lett.}\ }\textbf {\bibinfo {volume} {99}},\ \bibinfo {pages} {221914} (\bibinfo {year} {2011})}\BibitemShut {NoStop}%
\bibitem [{\citenamefont {Solovyev}(2025)}]{condmat10020021}%
  \BibitemOpen
  \bibfield  {author} {\bibinfo {author} {\bibfnamefont {I.~V.}\ \bibnamefont {Solovyev}},\ }\bibfield  {title} {\bibinfo {title} {Basic aspects of ferroelectricity induced by noncollinear alignment of spins},\ }\href {https://www.mdpi.com/2410-3896/10/2/21} {\bibfield  {journal} {\bibinfo  {journal} {Condensed Matter}\ }\textbf {\bibinfo {volume} {10}} (\bibinfo {year} {2025})}\BibitemShut {NoStop}%
\bibitem [{\citenamefont {Andreev}(2019)}]{Andreev_2019}%
  \BibitemOpen
  \bibfield  {author} {\bibinfo {author} {\bibfnamefont {P.~A.}\ \bibnamefont {Andreev}},\ }\bibfield  {title} {\bibinfo {title} {Hydrodynamic model of a $\mathrm{B}$ose–$\mathrm{E}$instein condensate with anisotropic short-range interaction and bright solitons in a repulsive $\mathrm{B}$ose–$\mathrm{E}$instein condensate},\ }\href {https://doi.org/10.1088/1555-6611/aaf921} {\bibfield  {journal} {\bibinfo  {journal} {Laser Physics}\ }\textbf {\bibinfo {volume} {29}},\ \bibinfo {pages} {035502} (\bibinfo {year} {2019})}\BibitemShut {NoStop}%
\bibitem [{\citenamefont {Andreev}\ and\ \citenamefont {Kuz'menkov}(2015)}]{Andreev_2016}%
  \BibitemOpen
  \bibfield  {author} {\bibinfo {author} {\bibfnamefont {P.~A.}\ \bibnamefont {Andreev}}\ and\ \bibinfo {author} {\bibfnamefont {L.~S.}\ \bibnamefont {Kuz'menkov}},\ }\bibfield  {title} {\bibinfo {title} {Separated spin-up and spin-down evolution of degenerated electrons in two-dimensional systems: Dispersion of longitudinal collective excitations in plane and nanotube geometry},\ }\href {https://doi.org/10.1209/0295-5075/113/17001} {\bibfield  {journal} {\bibinfo  {journal} {Europhys. Lett.}\ }\textbf {\bibinfo {volume} {113}},\ \bibinfo {pages} {17001} (\bibinfo {year} {2015})}\BibitemShut {NoStop}%
\bibitem [{\citenamefont {Andreev}\ and\ \citenamefont {Trukhanova}(2023)}]{Andreev_2023}%
  \BibitemOpen
  \bibfield  {author} {\bibinfo {author} {\bibfnamefont {P.~A.}\ \bibnamefont {Andreev}}\ and\ \bibinfo {author} {\bibfnamefont {M.~I.}\ \bibnamefont {Trukhanova}},\ }\bibfield  {title} {\bibinfo {title} {Quantum hydrodynamic representation of the exchange interaction in the theory of description of magnetically ordered media},\ }\href {https://doi.org/10.3103/S0027134923040021} {\bibfield  {journal} {\bibinfo  {journal} {Moscow Univ. Phys.}\ }\textbf {\bibinfo {volume} {78}},\ \bibinfo {pages} {445} (\bibinfo {year} {2023})}\BibitemShut {NoStop}%
\bibitem [{\citenamefont {Andreev}\ and\ \citenamefont {Trukhanova}(2024{\natexlab{d}})}]{JETP}%
  \BibitemOpen
  \bibfield  {author} {\bibinfo {author} {\bibfnamefont {P.~A.}\ \bibnamefont {Andreev}}\ and\ \bibinfo {author} {\bibfnamefont {M.~I.}\ \bibnamefont {Trukhanova}},\ }\bibfield  {title} {\bibinfo {title} {The evolution equation of the electric polarization of multiferroics, proportional to the vector product of the spins of the ions of the cell, under the influence of the heisenberg hamiltonian},\ }\href {https://doi.org/10.31857/S0044451024110099} {\bibfield  {journal} {\bibinfo  {journal} {JETP}\ }\textbf {\bibinfo {volume} {166}},\ \bibinfo {pages} {665} (\bibinfo {year} {2024}{\natexlab{d}})}\BibitemShut {NoStop}%
\bibitem [{\citenamefont {Dong}\ \emph {et~al.}(2019)\citenamefont {Dong}, \citenamefont {Xiang},\ and\ \citenamefont {Dagotto}}]{10.1093/nsr/nwz023}%
  \BibitemOpen
  \bibfield  {author} {\bibinfo {author} {\bibfnamefont {S.}~\bibnamefont {Dong}}, \bibinfo {author} {\bibfnamefont {H.}~\bibnamefont {Xiang}},\ and\ \bibinfo {author} {\bibfnamefont {E.}~\bibnamefont {Dagotto}},\ }\bibfield  {title} {\bibinfo {title} {Magnetoelectricity in multiferroics: a theoretical perspective},\ }\href {https://doi.org/10.1093/nsr/nwz023} {\bibfield  {journal} {\bibinfo  {journal} {National Science Review}\ }\textbf {\bibinfo {volume} {6}},\ \bibinfo {pages} {629} (\bibinfo {year} {2019})},\ \Eprint {https://arxiv.org/abs/https://academic.oup.com/nsr/article-pdf/6/4/629/38916012/nwz023.pdf} {https://academic.oup.com/nsr/article-pdf/6/4/629/38916012/nwz023.pdf} \BibitemShut {NoStop}%
\bibitem [{\citenamefont {Jensen}(1977)}]{JENSEN197732}%
  \BibitemOpen
  \bibfield  {author} {\bibinfo {author} {\bibfnamefont {J.}~\bibnamefont {Jensen}},\ }\bibfield  {title} {\bibinfo {title} {Two- and single-ion magnetic anisotropy in the rare-earth metals},\ }\href {https://doi.org/https://doi.org/10.1016/0378-4363(77)90215-7} {\bibfield  {journal} {\bibinfo  {journal} {Physica B+C}\ }\textbf {\bibinfo {volume} {86-88}},\ \bibinfo {pages} {32} (\bibinfo {year} {1977})}\BibitemShut {NoStop}%
\bibitem [{\citenamefont {Ganyushin}\ and\ \citenamefont {Neese}(2013)}]{Ganyushin2013}%
  \BibitemOpen
  \bibfield  {author} {\bibinfo {author} {\bibfnamefont {D.}~\bibnamefont {Ganyushin}}\ and\ \bibinfo {author} {\bibfnamefont {F.}~\bibnamefont {Neese}},\ }\bibfield  {title} {\bibinfo {title} {A fully variational spin-orbit coupled complete active space self-consistent field approach: Application to electron paramagnetic resonance g-tensors},\ }\href {https://doi.org/10.1063/1.4793736} {\bibfield  {journal} {\bibinfo  {journal} {The Journal of Chemical Physics}\ }\textbf {\bibinfo {volume} {138}},\ \bibinfo {pages} {104113} (\bibinfo {year} {2013})}\BibitemShut {NoStop}%
\bibitem [{\citenamefont {Rudowicz}\ and\ \citenamefont {Karbowiak}(2015)}]{RUDOWICZ201528}%
  \BibitemOpen
  \bibfield  {author} {\bibinfo {author} {\bibfnamefont {C.}~\bibnamefont {Rudowicz}}\ and\ \bibinfo {author} {\bibfnamefont {M.}~\bibnamefont {Karbowiak}},\ }\bibfield  {title} {\bibinfo {title} {Disentangling intricate web of interrelated notions at the interface between the physical (crystal field) hamiltonians and the effective (spin) hamiltonians},\ }\href {https://doi.org/https://doi.org/10.1016/j.ccr.2014.12.006} {\bibfield  {journal} {\bibinfo  {journal} {Coordination Chemistry Reviews}\ }\textbf {\bibinfo {volume} {287}},\ \bibinfo {pages} {28} (\bibinfo {year} {2015})}\BibitemShut {NoStop}%
\bibitem [{\citenamefont {Podlesnyak}\ \emph {et~al.}(2021)\citenamefont {Podlesnyak}, \citenamefont {Nikitin},\ and\ \citenamefont {Ehlers}}]{Podlesnyak_2021}%
  \BibitemOpen
  \bibfield  {author} {\bibinfo {author} {\bibfnamefont {A.}~\bibnamefont {Podlesnyak}}, \bibinfo {author} {\bibfnamefont {S.~E.}\ \bibnamefont {Nikitin}},\ and\ \bibinfo {author} {\bibfnamefont {G.}~\bibnamefont {Ehlers}},\ }\bibfield  {title} {\bibinfo {title} {Low-energy spin dynamics in rare-earth perovskite oxides},\ }\href {https://doi.org/10.1088/1361-648X/ac1367} {\bibfield  {journal} {\bibinfo  {journal} {Journal of Physics: Condensed Matter}\ }\textbf {\bibinfo {volume} {33}},\ \bibinfo {pages} {403001} (\bibinfo {year} {2021})}\BibitemShut {NoStop}%
\bibitem [{\citenamefont {Sushkov}\ \emph {et~al.}(2019)\citenamefont {Sushkov}, \citenamefont {Choi},\ and\ \citenamefont {Cheong}}]{Sushkov2019}%
  \BibitemOpen
  \bibfield  {author} {\bibinfo {author} {\bibfnamefont {A.~B.}\ \bibnamefont {Sushkov}}, \bibinfo {author} {\bibfnamefont {Y.~J.}\ \bibnamefont {Choi}},\ and\ \bibinfo {author} {\bibfnamefont {S.-W.}\ \bibnamefont {Cheong}},\ }\bibfield  {title} {\bibinfo {title} {Interplay between spin dynamics and crystal field in multiferroic homno$_3$},\ }\href {https://doi.org/10.1103/PhysRevB.100.014427} {\bibfield  {journal} {\bibinfo  {journal} {Physical Review B}\ }\textbf {\bibinfo {volume} {100}},\ \bibinfo {pages} {014427} (\bibinfo {year} {2019})}\BibitemShut {NoStop}%
\bibitem [{\citenamefont {Bonville}\ \emph {et~al.}(2014)\citenamefont {Bonville}, \citenamefont {Hodges}, \citenamefont {P\'erez},\ and\ \citenamefont {Fauth}}]{Bonville2014}%
  \BibitemOpen
  \bibfield  {author} {\bibinfo {author} {\bibfnamefont {P.}~\bibnamefont {Bonville}}, \bibinfo {author} {\bibfnamefont {J.~A.}\ \bibnamefont {Hodges}}, \bibinfo {author} {\bibfnamefont {O.}~\bibnamefont {P\'erez}},\ and\ \bibinfo {author} {\bibfnamefont {F.}~\bibnamefont {Fauth}},\ }\bibfield  {title} {\bibinfo {title} {Crystal field and magnetism with wannier functions: Orthorhombic rare-earth manganites},\ }\href {https://doi.org/10.1016/j.jmmm.2014.01.076} {\bibfield  {journal} {\bibinfo  {journal} {Journal of Magnetism and Magnetic Materials}\ }\textbf {\bibinfo {volume} {358--359}},\ \bibinfo {pages} {228} (\bibinfo {year} {2014})}\BibitemShut {NoStop}%
\bibitem [{\citenamefont {Yang}\ \emph {et~al.}(2023)\citenamefont {Yang}, \citenamefont {Liang},\ and\ \citenamefont {Cui}}]{Yang}%
  \BibitemOpen
  \bibfield  {author} {\bibinfo {author} {\bibfnamefont {H.}~\bibnamefont {Yang}}, \bibinfo {author} {\bibfnamefont {J.}~\bibnamefont {Liang}},\ and\ \bibinfo {author} {\bibfnamefont {Q.}~\bibnamefont {Cui}},\ }\bibfield  {title} {\bibinfo {title} {First-principles calculations for dzyaloshinskii–moriya interaction},\ }\href {https://doi.org/10.1038/s42254-022-00529-0} {\bibfield  {journal} {\bibinfo  {journal} {Nat. Rev. Phys.}\ }\textbf {\bibinfo {volume} {5}},\ \bibinfo {pages} {43} (\bibinfo {year} {2023})}\BibitemShut {NoStop}%
\bibitem [{\citenamefont {Witte}\ \emph {et~al.}(2020)\citenamefont {Witte}, \citenamefont {Sarkar}, \citenamefont {Velasco}, \citenamefont {Kruk}, \citenamefont {Brand}, \citenamefont {Eggert}, \citenamefont {Ollefs}, \citenamefont {Weschke}, \citenamefont {Wende},\ and\ \citenamefont {Hahn}}]{10.1063/5.0004125}%
  \BibitemOpen
  \bibfield  {author} {\bibinfo {author} {\bibfnamefont {R.}~\bibnamefont {Witte}}, \bibinfo {author} {\bibfnamefont {A.}~\bibnamefont {Sarkar}}, \bibinfo {author} {\bibfnamefont {L.}~\bibnamefont {Velasco}}, \bibinfo {author} {\bibfnamefont {R.}~\bibnamefont {Kruk}}, \bibinfo {author} {\bibfnamefont {R.~A.}\ \bibnamefont {Brand}}, \bibinfo {author} {\bibfnamefont {B.}~\bibnamefont {Eggert}}, \bibinfo {author} {\bibfnamefont {K.}~\bibnamefont {Ollefs}}, \bibinfo {author} {\bibfnamefont {E.}~\bibnamefont {Weschke}}, \bibinfo {author} {\bibfnamefont {H.}~\bibnamefont {Wende}},\ and\ \bibinfo {author} {\bibfnamefont {H.}~\bibnamefont {Hahn}},\ }\bibfield  {title} {\bibinfo {title} {Magnetic properties of rare-earth and transition metal based perovskite type high entropy oxides},\ }\href {https://doi.org/10.1063/5.0004125} {\bibfield  {journal} {\bibinfo  {journal} {Journal of Applied Physics}\ }\textbf {\bibinfo {volume} {127}},\ \bibinfo {pages} {185109} (\bibinfo {year} {2020})}\BibitemShut {NoStop}%
\bibitem [{\citenamefont {Giovannetti}\ \emph {et~al.}(2009)\citenamefont {Giovannetti}, \citenamefont {Kumar}, \citenamefont {Khomskii}, \citenamefont {Picozzi},\ and\ \citenamefont {van~den Brink}}]{PhysRevLett.103.156401}%
  \BibitemOpen
  \bibfield  {author} {\bibinfo {author} {\bibfnamefont {G.}~\bibnamefont {Giovannetti}}, \bibinfo {author} {\bibfnamefont {S.}~\bibnamefont {Kumar}}, \bibinfo {author} {\bibfnamefont {D.}~\bibnamefont {Khomskii}}, \bibinfo {author} {\bibfnamefont {S.}~\bibnamefont {Picozzi}},\ and\ \bibinfo {author} {\bibfnamefont {J.}~\bibnamefont {van~den Brink}},\ }\bibfield  {title} {\bibinfo {title} {Multiferroicity in rare-earth nickelates $r{\mathrm{nio}}_{3}$},\ }\href {https://doi.org/10.1103/PhysRevLett.103.156401} {\bibfield  {journal} {\bibinfo  {journal} {Phys. Rev. Lett.}\ }\textbf {\bibinfo {volume} {103}},\ \bibinfo {pages} {156401} (\bibinfo {year} {2009})}\BibitemShut {NoStop}%
\bibitem [{\citenamefont {Scaramucci}\ \emph {et~al.}(2012)\citenamefont {Scaramucci}, \citenamefont {Bousquet}, \citenamefont {Fechner}, \citenamefont {Mostovoy},\ and\ \citenamefont {Spaldin}}]{PhysRevLett.109.197203}%
  \BibitemOpen
  \bibfield  {author} {\bibinfo {author} {\bibfnamefont {A.}~\bibnamefont {Scaramucci}}, \bibinfo {author} {\bibfnamefont {E.}~\bibnamefont {Bousquet}}, \bibinfo {author} {\bibfnamefont {M.}~\bibnamefont {Fechner}}, \bibinfo {author} {\bibfnamefont {M.}~\bibnamefont {Mostovoy}},\ and\ \bibinfo {author} {\bibfnamefont {N.~A.}\ \bibnamefont {Spaldin}},\ }\bibfield  {title} {\bibinfo {title} {Linear magnetoelectric effect by orbital magnetism},\ }\href {https://doi.org/10.1103/PhysRevLett.109.197203} {\bibfield  {journal} {\bibinfo  {journal} {Phys. Rev. Lett.}\ }\textbf {\bibinfo {volume} {109}},\ \bibinfo {pages} {197203} (\bibinfo {year} {2012})}\BibitemShut {NoStop}%
\bibitem [{\citenamefont {Kajimoto}\ \emph {et~al.}(2011)\citenamefont {Kajimoto}, \citenamefont {Yoshizawa}, \citenamefont {Shintani}, \citenamefont {Kimura},\ and\ \citenamefont {Tokura}}]{Kajimoto2011}%
  \BibitemOpen
  \bibfield  {author} {\bibinfo {author} {\bibfnamefont {R.}~\bibnamefont {Kajimoto}}, \bibinfo {author} {\bibfnamefont {H.}~\bibnamefont {Yoshizawa}}, \bibinfo {author} {\bibfnamefont {H.}~\bibnamefont {Shintani}}, \bibinfo {author} {\bibfnamefont {T.}~\bibnamefont {Kimura}},\ and\ \bibinfo {author} {\bibfnamefont {Y.}~\bibnamefont {Tokura}},\ }\bibfield  {title} {\bibinfo {title} {Magnetic structure and ferroelectricity in multiferroic tbmno$_3$},\ }\href {https://doi.org/10.1103/PhysRevB.83.144402} {\bibfield  {journal} {\bibinfo  {journal} {Physical Review B}\ }\textbf {\bibinfo {volume} {83}},\ \bibinfo {pages} {144402} (\bibinfo {year} {2011})}\BibitemShut {NoStop}%
\bibitem [{\citenamefont {Soumya}\ \emph {et~al.}(2023)\citenamefont {Soumya}, \citenamefont {Vinod}, \citenamefont {Harsita}, \citenamefont {Sreelatha}, \citenamefont {{Durga Rao}}, \citenamefont {{Ramesh Kumar}}, \citenamefont {Rout}, \citenamefont {Gangopadhyay}, \citenamefont {Bhatnagar},\ and\ \citenamefont {Sattibabu}}]{SOUMYA2023123971}%
  \BibitemOpen
  \bibfield  {author} {\bibinfo {author} {\bibfnamefont {S.}~\bibnamefont {Soumya}}, \bibinfo {author} {\bibfnamefont {K.}~\bibnamefont {Vinod}}, \bibinfo {author} {\bibfnamefont {M.}~\bibnamefont {Harsita}}, \bibinfo {author} {\bibfnamefont {K.}~\bibnamefont {Sreelatha}}, \bibinfo {author} {\bibfnamefont {T.}~\bibnamefont {{Durga Rao}}}, \bibinfo {author} {\bibfnamefont {K.}~\bibnamefont {{Ramesh Kumar}}}, \bibinfo {author} {\bibfnamefont {J.}~\bibnamefont {Rout}}, \bibinfo {author} {\bibfnamefont {P.}~\bibnamefont {Gangopadhyay}}, \bibinfo {author} {\bibfnamefont {A.}~\bibnamefont {Bhatnagar}},\ and\ \bibinfo {author} {\bibfnamefont {B.}~\bibnamefont {Sattibabu}},\ }\bibfield  {title} {\bibinfo {title} {Studies on the effect of $\mathrm{I}$n$^{3+}$ ion on magnetic and magneto caloric properties of polycrystalline $\mathrm{T}$b$\mathrm{M}$n$\mathrm{O}_3$},\ }\href {https://doi.org/https://doi.org/10.1016/j.jssc.2023.123971} {\bibfield  {journal} {\bibinfo  {journal} {Journal of Solid State Chemistry}\
  }\textbf {\bibinfo {volume} {322}},\ \bibinfo {pages} {123971} (\bibinfo {year} {2023})}\BibitemShut {NoStop}%
\bibitem [{\citenamefont {Mansouri}\ \emph {et~al.}(2018)\citenamefont {Mansouri}, \citenamefont {Jandl}, \citenamefont {Balli}, \citenamefont {Fournier}, \citenamefont {Mukhin}, \citenamefont {Ivanov}, \citenamefont {Balbashov},\ and\ \citenamefont {Orlita}}]{Mansouri2018}%
  \BibitemOpen
  \bibfield  {author} {\bibinfo {author} {\bibfnamefont {S.}~\bibnamefont {Mansouri}}, \bibinfo {author} {\bibfnamefont {S.}~\bibnamefont {Jandl}}, \bibinfo {author} {\bibfnamefont {M.}~\bibnamefont {Balli}}, \bibinfo {author} {\bibfnamefont {P.}~\bibnamefont {Fournier}}, \bibinfo {author} {\bibfnamefont {A.~A.}\ \bibnamefont {Mukhin}}, \bibinfo {author} {\bibfnamefont {V.~Y.}\ \bibnamefont {Ivanov}}, \bibinfo {author} {\bibfnamefont {A.}~\bibnamefont {Balbashov}},\ and\ \bibinfo {author} {\bibfnamefont {M.}~\bibnamefont {Orlita}},\ }\bibfield  {title} {\bibinfo {title} {Study of crystal-field excitations and infrared active phonons in tbmno$_3$},\ }\href {https://doi.org/10.1088/1361-648X/aaaf06} {\bibfield  {journal} {\bibinfo  {journal} {Journal of Physics: Condensed Matter}\ }\textbf {\bibinfo {volume} {30}},\ \bibinfo {pages} {175602} (\bibinfo {year} {2018})}\BibitemShut {NoStop}%
\bibitem [{\citenamefont {Kashchenko}\ \emph {et~al.}(2017)\citenamefont {Kashchenko}, \citenamefont {Klimin}, \citenamefont {Balbashov},\ and\ \citenamefont {Popova}}]{Kashchenko2017}%
  \BibitemOpen
  \bibfield  {author} {\bibinfo {author} {\bibfnamefont {M.~A.}\ \bibnamefont {Kashchenko}}, \bibinfo {author} {\bibfnamefont {S.~A.}\ \bibnamefont {Klimin}}, \bibinfo {author} {\bibfnamefont {A.~M.}\ \bibnamefont {Balbashov}},\ and\ \bibinfo {author} {\bibfnamefont {M.~N.}\ \bibnamefont {Popova}},\ }\bibfield  {title} {\bibinfo {title} {Probing dy$^{3+}$ magnetic moments in multiferroic perovskite dymno$_3$ by optical spectroscopy},\ }\href {https://doi.org/10.1103/PhysRevB.95.174422} {\bibfield  {journal} {\bibinfo  {journal} {Physical Review B}\ }\textbf {\bibinfo {volume} {95}},\ \bibinfo {pages} {174422} (\bibinfo {year} {2017})}\BibitemShut {NoStop}%
\bibitem [{\citenamefont {Dubrovskiy}\ \emph {et~al.}(2017)\citenamefont {Dubrovskiy}, \citenamefont {Rautskii}, \citenamefont {Moshkina},\ and\ \citenamefont {et~al.}}]{Dubrovskiy}%
  \BibitemOpen
  \bibfield  {author} {\bibinfo {author} {\bibfnamefont {A.}~\bibnamefont {Dubrovskiy}}, \bibinfo {author} {\bibfnamefont {M.}~\bibnamefont {Rautskii}}, \bibinfo {author} {\bibfnamefont {E.}~\bibnamefont {Moshkina}},\ and\ \bibinfo {author} {\bibnamefont {et~al.}},\ }\bibfield  {title} {\bibinfo {title} {$\mathrm{EPR}$-determined anisotropy of the g-factor and magnetostriction of a $\mathrm{C}$u$_2$$\mathrm{M}$n$\mathrm{B}$$\mathrm{O}_5$ single crystal with a ludwigite structure},\ }\href {https://doi.org/10.1134/S0021364017230072} {\bibfield  {journal} {\bibinfo  {journal} {Jetp Lett.}\ }\textbf {\bibinfo {volume} {106}},\ \bibinfo {pages} {716} (\bibinfo {year} {2017})}\BibitemShut {NoStop}%
\bibitem [{\citenamefont {Sagayama}\ \emph {et~al.}(2008)\citenamefont {Sagayama}, \citenamefont {Taniguchi}, \citenamefont {Abe}, \citenamefont {Arima}, \citenamefont {Soda}, \citenamefont {Matsuura},\ and\ \citenamefont {Hirota}}]{PhysRevB.77.220407}%
  \BibitemOpen
  \bibfield  {author} {\bibinfo {author} {\bibfnamefont {H.}~\bibnamefont {Sagayama}}, \bibinfo {author} {\bibfnamefont {K.}~\bibnamefont {Taniguchi}}, \bibinfo {author} {\bibfnamefont {N.}~\bibnamefont {Abe}}, \bibinfo {author} {\bibfnamefont {T.-h.}\ \bibnamefont {Arima}}, \bibinfo {author} {\bibfnamefont {M.}~\bibnamefont {Soda}}, \bibinfo {author} {\bibfnamefont {M.}~\bibnamefont {Matsuura}},\ and\ \bibinfo {author} {\bibfnamefont {K.}~\bibnamefont {Hirota}},\ }\bibfield  {title} {\bibinfo {title} {Correlation between ferroelectric polarization and sense of helical spin order in multiferroic ${\text{mnwo}}_{4}$},\ }\href {https://doi.org/10.1103/PhysRevB.77.220407} {\bibfield  {journal} {\bibinfo  {journal} {Phys. Rev. B}\ }\textbf {\bibinfo {volume} {77}},\ \bibinfo {pages} {220407(R)} (\bibinfo {year} {2008})}\BibitemShut {NoStop}%
\bibitem [{\citenamefont {Tolédano}\ \emph {et~al.}(2010)\citenamefont {Tolédano}, \citenamefont {Mettout}, \citenamefont {Schranz},\ and\ \citenamefont {Krexner}}]{Toledano_2010}%
  \BibitemOpen
  \bibfield  {author} {\bibinfo {author} {\bibfnamefont {P.}~\bibnamefont {Tolédano}}, \bibinfo {author} {\bibfnamefont {B.}~\bibnamefont {Mettout}}, \bibinfo {author} {\bibfnamefont {W.}~\bibnamefont {Schranz}},\ and\ \bibinfo {author} {\bibfnamefont {G.}~\bibnamefont {Krexner}},\ }\bibfield  {title} {\bibinfo {title} {Directional magnetoelectric effects in mnwo4: magnetic sources of the electric polarization},\ }\href {https://doi.org/10.1088/0953-8984/22/6/065901} {\bibfield  {journal} {\bibinfo  {journal} {Journal of Physics: Condensed Matter}\ }\textbf {\bibinfo {volume} {22}},\ \bibinfo {pages} {065901} (\bibinfo {year} {2010})}\BibitemShut {NoStop}%
\bibitem [{\citenamefont {Khan}\ \emph {et~al.}(2020)\citenamefont {Khan}, \citenamefont {Bonville},\ and\ \citenamefont {Hodges}}]{Khan2020}%
  \BibitemOpen
  \bibfield  {author} {\bibinfo {author} {\bibfnamefont {M.~A.}\ \bibnamefont {Khan}}, \bibinfo {author} {\bibfnamefont {P.}~\bibnamefont {Bonville}},\ and\ \bibinfo {author} {\bibfnamefont {J.~A.}\ \bibnamefont {Hodges}},\ }\bibfield  {title} {\bibinfo {title} {Crystal field and magnetic anisotropy in rare-earth orthoferrites},\ }\href {https://doi.org/10.1103/PhysRevB.102.174419} {\bibfield  {journal} {\bibinfo  {journal} {Physical Review B}\ }\textbf {\bibinfo {volume} {102}},\ \bibinfo {pages} {174419} (\bibinfo {year} {2020})}\BibitemShut {NoStop}%
\bibitem [{\citenamefont {Zorko}\ \emph {et~al.}(2011)\citenamefont {Zorko}, \citenamefont {Pregelj}, \citenamefont {Gomil\v{s}ek}, \citenamefont {van Tol}, \citenamefont {Ozarowski},\ and\ \citenamefont {Ar\v{c}on}}]{Zorko2011}%
  \BibitemOpen
  \bibfield  {author} {\bibinfo {author} {\bibfnamefont {A.}~\bibnamefont {Zorko}}, \bibinfo {author} {\bibfnamefont {M.}~\bibnamefont {Pregelj}}, \bibinfo {author} {\bibfnamefont {M.}~\bibnamefont {Gomil\v{s}ek}}, \bibinfo {author} {\bibfnamefont {J.}~\bibnamefont {van Tol}}, \bibinfo {author} {\bibfnamefont {A.}~\bibnamefont {Ozarowski}},\ and\ \bibinfo {author} {\bibfnamefont {D.}~\bibnamefont {Ar\v{c}on}},\ }\bibfield  {title} {\bibinfo {title} {Dzyaloshinsky-moriya interaction in the spin-1/2 heisenberg antiferromagnet on a pyrochlore lattice},\ }\href {https://doi.org/10.1103/PhysRevLett.107.257203} {\bibfield  {journal} {\bibinfo  {journal} {Physical Review Letters}\ }\textbf {\bibinfo {volume} {107}},\ \bibinfo {pages} {257203} (\bibinfo {year} {2011})}\BibitemShut {NoStop}%
\bibitem [{\citenamefont {Pradhan}\ \emph {et~al.}(2025)\citenamefont {Pradhan}, \citenamefont {Panda}, \citenamefont {Mishra},\ and\ \citenamefont {Mohanty}}]{PRADHAN2025100082}%
  \BibitemOpen
  \bibfield  {author} {\bibinfo {author} {\bibfnamefont {P.~K.}\ \bibnamefont {Pradhan}}, \bibinfo {author} {\bibfnamefont {A.}~\bibnamefont {Panda}}, \bibinfo {author} {\bibfnamefont {G.}~\bibnamefont {Mishra}},\ and\ \bibinfo {author} {\bibfnamefont {N.}~\bibnamefont {Mohanty}},\ }\bibfield  {title} {\bibinfo {title} {Rare earth orthoferrites (rfeo3, r= rare earth elements): A comprehensive review of structural, dielectric, and magnetic properties},\ }\href {https://doi.org/https://doi.org/10.1016/j.smmf.2025.100082} {\bibfield  {journal} {\bibinfo  {journal} {Smart Materials in Manufacturing}\ }\textbf {\bibinfo {volume} {3}},\ \bibinfo {pages} {100082} (\bibinfo {year} {2025})}\BibitemShut {NoStop}%
\bibitem [{\citenamefont {Song}\ \emph {et~al.}(2019)\citenamefont {Song}, \citenamefont {Xu},\ and\ \citenamefont {Nan}}]{10.1093/nsr/nwz055}%
  \BibitemOpen
  \bibfield  {author} {\bibinfo {author} {\bibfnamefont {Y.}~\bibnamefont {Song}}, \bibinfo {author} {\bibfnamefont {B.}~\bibnamefont {Xu}},\ and\ \bibinfo {author} {\bibfnamefont {C.-W.}\ \bibnamefont {Nan}},\ }\bibfield  {title} {\bibinfo {title} {Lattice and spin dynamics in multiferroic bifeo3 and rmno3},\ }\href {https://doi.org/10.1093/nsr/nwz055} {\bibfield  {journal} {\bibinfo  {journal} {National Science Review}\ }\textbf {\bibinfo {volume} {6}},\ \bibinfo {pages} {642} (\bibinfo {year} {2019})},\ \Eprint {https://arxiv.org/abs/https://academic.oup.com/nsr/article-pdf/6/4/642/38916046/nwz055.pdf} {https://academic.oup.com/nsr/article-pdf/6/4/642/38916046/nwz055.pdf} \BibitemShut {NoStop}%
\bibitem [{\citenamefont {Kimura}\ \emph {et~al.}(2003)\citenamefont {Kimura}, \citenamefont {Goto}, \citenamefont {Shintani},\ and\ \citenamefont {et~al.}}]{10.10631.3665634}%
  \BibitemOpen
  \bibfield  {author} {\bibinfo {author} {\bibfnamefont {T.}~\bibnamefont {Kimura}}, \bibinfo {author} {\bibfnamefont {T.}~\bibnamefont {Goto}}, \bibinfo {author} {\bibfnamefont {H.}~\bibnamefont {Shintani}},\ and\ \bibinfo {author} {\bibnamefont {et~al.}},\ }\bibfield  {title} {\bibinfo {title} {Magnetic control of ferroelectric polarization},\ }\href {https://doi.org/10.1038/nature02018} {\bibfield  {journal} {\bibinfo  {journal} {Nature}\ }\textbf {\bibinfo {volume} {426}},\ \bibinfo {pages} {55} (\bibinfo {year} {2003})}\BibitemShut {NoStop}%
\bibitem [{\citenamefont {Kimura}\ \emph {et~al.}(2007)\citenamefont {Kimura}, \citenamefont {Lawes},\ and\ \citenamefont {Ramirez}}]{Kimura2006}%
  \BibitemOpen
  \bibfield  {author} {\bibinfo {author} {\bibfnamefont {T.}~\bibnamefont {Kimura}}, \bibinfo {author} {\bibfnamefont {G.}~\bibnamefont {Lawes}},\ and\ \bibinfo {author} {\bibfnamefont {A.~P.}\ \bibnamefont {Ramirez}},\ }\bibfield  {title} {\bibinfo {title} {Electric polarization rotation in a hexaferrite with long-wavelength magnetic helices},\ }\href {https://doi.org/10.1103/PhysRevLett.98.187201} {\bibfield  {journal} {\bibinfo  {journal} {Physical Review Letters}\ }\textbf {\bibinfo {volume} {98}},\ \bibinfo {pages} {187201} (\bibinfo {year} {2007})}\BibitemShut {NoStop}%
\bibitem [{\citenamefont {Johnson}\ \emph {et~al.}(2012)\citenamefont {Johnson}, \citenamefont {Chapon}, \citenamefont {Khalyavin}, \citenamefont {Manuel},\ and\ \citenamefont {Radaelli}}]{Johnson2012}%
  \BibitemOpen
  \bibfield  {author} {\bibinfo {author} {\bibfnamefont {R.~D.}\ \bibnamefont {Johnson}}, \bibinfo {author} {\bibfnamefont {L.~C.}\ \bibnamefont {Chapon}}, \bibinfo {author} {\bibfnamefont {D.~D.}\ \bibnamefont {Khalyavin}}, \bibinfo {author} {\bibfnamefont {P.}~\bibnamefont {Manuel}},\ and\ \bibinfo {author} {\bibfnamefont {P.~G.}\ \bibnamefont {Radaelli}},\ }\bibfield  {title} {\bibinfo {title} {Magnetic order and magnetoelectric coupling in the proper-screw multiferroic cucro$_2$},\ }\href {https://doi.org/10.1103/PhysRevLett.108.067201} {\bibfield  {journal} {\bibinfo  {journal} {Physical Review Letters}\ }\textbf {\bibinfo {volume} {108}},\ \bibinfo {pages} {067201} (\bibinfo {year} {2012})}\BibitemShut {NoStop}%
\bibitem [{\citenamefont {Andreev}(2025)}]{andreev2025keffer}%
  \BibitemOpen
  \bibfield  {author} {\bibinfo {author} {\bibfnamefont {P.~A.}\ \bibnamefont {Andreev}},\ }\bibfield  {title} {\bibinfo {title} {Keffer-like form of the symmetric heisenberg exchange integral: Contribution to the $\mathrm{L}$andau-$\mathrm{L}$ifshitz-$\mathrm{G}$ilbert equation and spin wave dispersion dependence},\ }\href {https://arxiv.org/abs/2512.22108} {\bibfield  {journal} {\bibinfo  {journal} {2512.22108}\ } (\bibinfo {year} {2025})}\BibitemShut {NoStop}%
\bibitem [{\citenamefont {Sergienko}\ and\ \citenamefont {Dagotto}(2006{\natexlab{b}})}]{Sergienko2006}%
  \BibitemOpen
  \bibfield  {author} {\bibinfo {author} {\bibfnamefont {I.~A.}\ \bibnamefont {Sergienko}}\ and\ \bibinfo {author} {\bibfnamefont {E.}~\bibnamefont {Dagotto}},\ }\bibfield  {title} {\bibinfo {title} {Role of the dzyaloshinskii--moriya interaction in multiferroic perovskites},\ }\href {https://doi.org/10.1103/PhysRevB.73.094434} {\bibfield  {journal} {\bibinfo  {journal} {Physical Review B}\ }\textbf {\bibinfo {volume} {73}},\ \bibinfo {pages} {094434} (\bibinfo {year} {2006}{\natexlab{b}})}\BibitemShut {NoStop}%
\bibitem [{\citenamefont {Rubins}\ \emph {et~al.}(2014)\citenamefont {Rubins}, \citenamefont {Popov}, \citenamefont {Kuzmin},\ and\ \citenamefont {Kotlov}}]{Rubins2014}%
  \BibitemOpen
  \bibfield  {author} {\bibinfo {author} {\bibfnamefont {R.}~\bibnamefont {Rubins}}, \bibinfo {author} {\bibfnamefont {A.~I.}\ \bibnamefont {Popov}}, \bibinfo {author} {\bibfnamefont {A.}~\bibnamefont {Kuzmin}},\ and\ \bibinfo {author} {\bibfnamefont {A.}~\bibnamefont {Kotlov}},\ }\bibfield  {title} {\bibinfo {title} {Electron paramagnetic resonance and crystal-field analysis of rare-earth ions in low-symmetry crystals},\ }\href {https://doi.org/10.1063/1.4887379} {\bibfield  {journal} {\bibinfo  {journal} {Journal of Applied Physics}\ }\textbf {\bibinfo {volume} {116}},\ \bibinfo {pages} {023501} (\bibinfo {year} {2014})}\BibitemShut {NoStop}%
\bibitem [{\citenamefont {Finger}\ \emph {et~al.}(2014)\citenamefont {Finger}, \citenamefont {Binder}, \citenamefont {Sidis}, \citenamefont {Maljuk}, \citenamefont {Argyriou},\ and\ \citenamefont {Braden}}]{PhysRevB.90.224418}%
  \BibitemOpen
  \bibfield  {author} {\bibinfo {author} {\bibfnamefont {T.}~\bibnamefont {Finger}}, \bibinfo {author} {\bibfnamefont {K.}~\bibnamefont {Binder}}, \bibinfo {author} {\bibfnamefont {Y.}~\bibnamefont {Sidis}}, \bibinfo {author} {\bibfnamefont {A.}~\bibnamefont {Maljuk}}, \bibinfo {author} {\bibfnamefont {D.~N.}\ \bibnamefont {Argyriou}},\ and\ \bibinfo {author} {\bibfnamefont {M.}~\bibnamefont {Braden}},\ }\bibfield  {title} {\bibinfo {title} {Magnetic order and electromagnon excitations in ${\mathrm{dymno}}_{3}$ studied by neutron scattering experiments},\ }\href {https://doi.org/10.1103/PhysRevB.90.224418} {\bibfield  {journal} {\bibinfo  {journal} {Phys. Rev. B}\ }\textbf {\bibinfo {volume} {90}},\ \bibinfo {pages} {224418} (\bibinfo {year} {2014})}\BibitemShut {NoStop}%
\bibitem [{\citenamefont {Lee}\ \emph {et~al.}(2021)\citenamefont {Lee}, \citenamefont {Huang}, \citenamefont {Jeng}, \citenamefont {Shao}, \citenamefont {Wray}, \citenamefont {Chen}, \citenamefont {Qiao}, \citenamefont {Yang}, \citenamefont {Lin}, \citenamefont {Schoenlein},\ and\ \citenamefont {Chuang}}]{LEE2021147013}%
  \BibitemOpen
  \bibfield  {author} {\bibinfo {author} {\bibfnamefont {J.-M.}\ \bibnamefont {Lee}}, \bibinfo {author} {\bibfnamefont {S.-W.}\ \bibnamefont {Huang}}, \bibinfo {author} {\bibfnamefont {H.-T.}\ \bibnamefont {Jeng}}, \bibinfo {author} {\bibfnamefont {Y.-C.}\ \bibnamefont {Shao}}, \bibinfo {author} {\bibfnamefont {L.~A.}\ \bibnamefont {Wray}}, \bibinfo {author} {\bibfnamefont {J.~M.}\ \bibnamefont {Chen}}, \bibinfo {author} {\bibfnamefont {R.}~\bibnamefont {Qiao}}, \bibinfo {author} {\bibfnamefont {W.}~\bibnamefont {Yang}}, \bibinfo {author} {\bibfnamefont {J.-Y.}\ \bibnamefont {Lin}}, \bibinfo {author} {\bibfnamefont {R.~W.}\ \bibnamefont {Schoenlein}},\ and\ \bibinfo {author} {\bibfnamefont {Y.-D.}\ \bibnamefont {Chuang}},\ }\bibfield  {title} {\bibinfo {title} {The magnetic order in multiferroic dymno3},\ }\href {https://doi.org/https://doi.org/10.1016/j.elspec.2020.147013} {\bibfield  {journal} {\bibinfo  {journal} {Journal of Electron Spectroscopy and Related Phenomena}\ }\textbf {\bibinfo {volume} {246}},\
  \bibinfo {pages} {147013} (\bibinfo {year} {2021})}\BibitemShut {NoStop}%
\bibitem [{\citenamefont {Heyer}\ \emph {et~al.}(2006)\citenamefont {Heyer}, \citenamefont {Hollmann}, \citenamefont {Klassen}, \citenamefont {Jodlauk}, \citenamefont {Bohatý}, \citenamefont {Becker}, \citenamefont {Mydosh}, \citenamefont {Lorenz},\ and\ \citenamefont {Khomskii}}]{Heyer_2006}%
  \BibitemOpen
  \bibfield  {author} {\bibinfo {author} {\bibfnamefont {O.}~\bibnamefont {Heyer}}, \bibinfo {author} {\bibfnamefont {N.}~\bibnamefont {Hollmann}}, \bibinfo {author} {\bibfnamefont {I.}~\bibnamefont {Klassen}}, \bibinfo {author} {\bibfnamefont {S.}~\bibnamefont {Jodlauk}}, \bibinfo {author} {\bibfnamefont {L.}~\bibnamefont {Bohatý}}, \bibinfo {author} {\bibfnamefont {P.}~\bibnamefont {Becker}}, \bibinfo {author} {\bibfnamefont {J.~A.}\ \bibnamefont {Mydosh}}, \bibinfo {author} {\bibfnamefont {T.}~\bibnamefont {Lorenz}},\ and\ \bibinfo {author} {\bibfnamefont {D.}~\bibnamefont {Khomskii}},\ }\bibfield  {title} {\bibinfo {title} {A new multiferroic material: Mnwo4},\ }\href {https://doi.org/10.1088/0953-8984/18/39/L01} {\bibfield  {journal} {\bibinfo  {journal} {Journal of Physics: Condensed Matter}\ }\textbf {\bibinfo {volume} {18}},\ \bibinfo {pages} {L471} (\bibinfo {year} {2006})}\BibitemShut {NoStop}%
\bibitem [{\citenamefont {Li}(2026)}]{LI2026173724}%
  \BibitemOpen
  \bibfield  {author} {\bibinfo {author} {\bibfnamefont {Y.}~\bibnamefont {Li}},\ }\bibfield  {title} {\bibinfo {title} {The lattice modulation and magnetic ferroelectric mechanism in multiferroics mnwo4},\ }\href {https://doi.org/https://doi.org/10.1016/j.jmmm.2025.173724} {\bibfield  {journal} {\bibinfo  {journal} {Journal of Magnetism and Magnetic Materials}\ }\textbf {\bibinfo {volume} {638}},\ \bibinfo {pages} {173724} (\bibinfo {year} {2026})}\BibitemShut {NoStop}%
\bibitem [{\citenamefont {Arima}(2007)}]{Arima2007}%
  \BibitemOpen
  \bibfield  {author} {\bibinfo {author} {\bibfnamefont {T.}~\bibnamefont {Arima}},\ }\bibfield  {title} {\bibinfo {title} {Ferroelectricity induced by proper-screw type magnetic order},\ }\href {https://doi.org/10.1143/JPSJ.76.073702} {\bibfield  {journal} {\bibinfo  {journal} {J. Phys. Soc. Jpn.}\ }\textbf {\bibinfo {volume} {76}},\ \bibinfo {pages} {073702} (\bibinfo {year} {2007})}\BibitemShut {NoStop}%
\end{thebibliography}
\end{document}